\documentclass[journal,peerreview,nodraftcls]{IEEEtran}
\usepackage{graphicx,subfig,amsthm,amsmath,latexsym,amssymb,times}
\usepackage{epsfig,multirow,rotating,times,verbatim,wrapfig,cite,booktabs,float}
\usepackage{color,xr,array,hyperref, algpseudocode}

\usepackage{array}

\usepackage{array,epsfig,rotating}
\usepackage[]{hyperref}  

\usepackage{graphicx}
\usepackage{float}
\usepackage{wrapfig}
\usepackage{color}

\usepackage{amsmath,verbatim}
\usepackage{tabularx}
\usepackage{amsmath}
\usepackage{amssymb}
\usepackage{amsfonts}
\usepackage{multirow}
\usepackage{amsthm}
\newcommand{\tr}{\mbox{tr}}
\newcommand{\abs}[1]{|#1|}

\newcommand{\R}{I\!R}

\newcommand{\btheta}{\boldsymbol{\theta}}

\newcommand{\bbeta}{\boldsymbol{\beta}}
\newcommand{\bseta}{\boldsymbol{\eta}}
\newcommand{\bdelta}{\boldsymbol{\delta}}

\newcommand{\bSigma}{\boldsymbol{\Sigma}}
\newcommand{\bsigma}{\boldsymbol{\sigma}}

\newcommand{\bepsilon}{\boldsymbol{\epsilon}}

\newcommand{\bmu}{\boldsymbol{\mu}}

\newcommand{\bB}{\boldsymbol{B}}
\newcommand{\bC}{\boldsymbol{C}}

\newcommand{\bg}{\boldsymbol{g}}
\newcommand{\bD}{\boldsymbol{D}}
\newcommand{\bff}{\boldsymbol{f}}
\newcommand{\bV}{\boldsymbol{V}}

\newcommand{\bh}{\boldsymbol{h}}

\newcommand{\bI}{\boldsymbol{I}}

\newcommand{\bK}{\boldsymbol{K}}

\newcommand{\bA}{\boldsymbol{A}}

\newcommand{\bS}{\boldsymbol{S}}

\newcommand{\bW}{\boldsymbol{W}}
\newcommand{\bu}{\boldsymbol{u}}

\newcommand{\bs}{\boldsymbol{s}}

\newcommand{\bx}{\boldsymbol{x}}
\newcommand{\bX}{\boldsymbol{X}}
\newcommand{\by}{\boldsymbol{y}}

\newcommand{\bzero}{\boldsymbol{0}}

\newcommand{\mD}{\mathcal D}

\newcommand{\citep}{\cite}
\newcommand{\citet}{\cite}
\newcommand{\citeyear}{\cite}
\newcommand{\citeauthor}{\cite}

\begin{document}
\title{Estimating Basis Functions in Massive Fields under the
 Spatial Mixed Effects Model}
\author{Karl T. Pazdernik~and~Ranjan~Maitra
  \thanks{K. T. Pazdernik is with the Pacific Northwest National
    Laboratories, Richland, Washington, USA.}
  \thanks{R.Maitra is with the Department of Statistics, Iowa State
    University, Ames, Iowa, USA.}
   \thanks{This research was supported, in part, by the National Science
     Foundation (NSF) under its CAREER Grant No. DMS-0437555 and the
     United States Department of Agriculture (USDA) National 
Institute of Food and Agriculture (NIFA) Hatch project IOW03617. 
The content of this paper is however solely the responsibility of the
authors and does not represent the official views of the NSF, the NIFA
or the USDA.}
\thanks{\copyright 2020 IEEE. Personal use of this material is
  permitted. However, permission to use this material for any other
  purposes must be obtained from the IEEE by sending a request to
  pubs-permissions@ieee.org.}
\thanks{Manuscript received xxxx xx,201x; revised xxxxxxxx xx,
  201x. Accepted xxxxxxxx xx, 201x.
  First published xxxxxxxx x, xxxx, current version published
  yyyyyyyy y, yyyy}
\thanks{Color versions of one or more of the figures in this paper are
  available online at http://ieeexplore.org.} 
\thanks{Digital Object Identifier}
}

\markboth{IEEE Transactions on Geosciences and Remote Sensing,~Vol.~?, No.~?, February~201?}%
{Pazdernik and Maitra: Estimating Basis Functions in Massive Fields}

\maketitle

\begin{abstract}

Spatial prediction is commonly achieved under the assumption of a Gaussian random  field (GRF) by obtaining maximum likelihood estimates of parameters, and then using the kriging equations to arrive at predicted values. For massive datasets, fixed rank kriging using the Expectation-Maximization (EM) algorithm for estimation has been proposed as an alternative to the usual but computationally prohibitive kriging method. The method reduces computation cost of estimation by redefining the spatial process as a linear combination of basis functions and spatial random effects. A disadvantage of this method is that it imposes constraints on the relationship between the observed locations and the knots. We develop an alternative  method that utilizes the Spatial Mixed Effects (SME) model, but allows for additional flexibility by estimating the range of the spatial dependence between the observations and the knots via an Alternating Expectation Conditional Maximization (AECM) algorithm. Experiments show that our methodology improves estimation without sacrificing prediction accuracy while also minimizing the additional computational burden of extra parameter estimation. The methodology is applied to a temperature data set archived by the United States National Climate Data Center, with improved results over previous methodology.
\end{abstract}


\section{Introduction}

In geostatistics, spatial estimation and prediction are often the primary 
focus. It is assumed that nearby observations tend to be more similar 
than those far apart. Since observations are spatially correlated, modeling 
the dependence structure can provide insight into the spatial phenomena and 
can be used to improve prediction. To formalize ideas, let $\bs =
\{\bs_1,\bs_2,\ldots,\bs_n\}\in \mathcal{D}$  
represent all observed locations and let $\bs_0 =
\{\bs_{01},\bs_{02},\ldots, \bs_{0N}\} \in \mathcal{D}$ represent the
desired locations, where $\mathcal{D}$ is  
a spatial domain. The two sets of locations $\bs$ and $\bs_0$ may (or may not) 
have common elements. Let $y(\cdot)$ be the random process defined on
$\mathcal{D}$. Generally, $y(\cdot)$ is characterized as a linear combination
of three main components: a mean structure $\mu(\cdot)$, a 
zero-mean spatial process $f(\cdot)$, and a zero-mean measurement 
error process $\epsilon(\cdot)$, {\em i.e.} 
\begin{equation}
\label{model1}
	\by(\bs) = \bmu(\bs) + \bff(\bs) + \bepsilon(\bs),
\end{equation}
where $f(\cdot)$ and $\epsilon(\cdot)$ are typically assumed to be
Gaussian processes. Any likelihood-based estimation procedure involves 
evaluating the  log-likelihood 
\begin{equation}
\label{logl}
\begin{split}
	\ell(\btheta) \propto & -\frac{1}{2}[\by(\bs) - \bmu(\bs)]'\{\mbox{Cov}[\by(\bs), 
	  \by(\bs)]\}^{-1}[\by(\bs) - \bmu(\bs)] \\
	  & - \frac{1}{2}\log\{|\mbox{Cov}[\by(\bs), \by(\bs)]|\}.
\end{split}
\end{equation}
Under squared error loss, the Best Linear Unbiased Predictions (BLUP)
-- equivalently, \cite{matheron1962}'s kriged estimate -- of $y(\cdot)$ at the
desired locations  $\bs_0$ can  
be computed using the first two moments of $y(\cdot)$ (see {\em e.g.} \cite{cressie1993}) as: 
\begin{equation}
\label{krige}
\begin{split}
	\hat\by(\bs_0) =& \mbox{E}(\by(\bs_0)) +
	\mbox{Cov}[\by(\bs_0), \by(\bs)]
	\{\mbox{Cov}[\by(\bs), \by(\bs)]\}^{-1} \\
	& [\by(\bs_0) - \mbox{E}(\by(\bs_0))].
\end{split}
\end{equation}
 
Both \eqref{logl} and \eqref{krige} involve inverting a covariance
matrix of order $n\times n$ which is an $O(n^3)$ operation and thus
computationally impractical for 
massive datasets in terms of both CPU time and memory. The need for
such predictions in massive spatial datasets however exist, for
example in satellite data where observations 
are recorded across the entire globe or more localized spatial domains 
in which the fine resolution results in a large $n$. In such scenarios, 
it is computationally infeasible to obtain the kriged estimates provided
by~\eqref{krige} even on modern computing hardware.

The development of efficient kriging methods has received substantial 
attention in the literature. Much of this development has focused
either on approximating the kriging equations of \eqref{krige} 
\cite{vecchia1988, nychkaetal1996, higdon1998, nychka2000,nychkawikleroyle2002, 
fuentes2002, billingsbeatsonnewsam2002a, billingsbeatsonnewsam2002b, 
steinetal2004, quinoerocandelarasmussen2005, furrergentonnychka2006, 
kaufmanschervishnychka2008} or on defining a covariance structure that
allows for exact kriging, regardless of the size of the data
\cite{huangcressiegabrosek2002, johannessoncressie2004, 
johannessoncressiehuang2007, cressiejohannesson2008, banerjeeetal2008, 
finleyetal2008, lindgren2011, nychka2015, katzfuss2016}.
In this paper, we develop methodology that is capable of modeling severe 
nonstationarity while providing useful parameter interpretation in an 
efficient manner and without loss of prediction accuracy. Our underlying 
framework uses the ``Spatial Mixed Effects'' (SME) model defined by 
\cite{cressiejohannesson2008} to obtain parameter estimates and 
spatial predictions, as it provides a good compromise between flexibility 
in the covariance structure and computational efficiency. 
See \cite{bradleyetal2014} for a more detailed comparison of some of the 
different approaches.

While \cite{finleyetal2008} allow for parameter estimation for nonstationary 
random fields, this methodology assumes the dependence structure between 
observations is the same as between observations and ``knots'', a 
lower-dimensional subspace. However, we desire a more flexible model that 
allows for cross-regional dependence, which can be achieved by the SME model. 
In addition, the computational cost is prohibitive for massive data, as pointed out 
by \cite{bradleyetal2014}. 
Although \cite{lindgren2011} and \cite{nychka2015} are equally as efficient, 
again, the flexibility of the assumed covariance structure is more restrictive. 

Use of so-called low-rank approaches, such as fixed rank kriging through the 
SME model, have been called into question by \cite{stein2014}. The author illustrates, 
through both theory and simulation, how prediction and model fit can actually be improved 
by assuming independent spatial block and using the corresponding methodology, 
such as \cite{vecchia1988,steinetal2004}. However, in this paper, our focus is on 
estimation and prediction of spatial models that may possess significant regional 
dependence, where the assumption of independent blocks is clearly inappropriate. 
Therefore, we now discuss the SME model in some detail.

\subsection{Spatial Mixed Effects Model}

Within the framework~\eqref{model1}, the SME 
model uses a linear combination of fixed basis functions and a set of 
locations with reduced cardinality ($m < n$) known as {\em knots} to define 
the spatial process at the original set $\bs$. A fine-scale 
variation process $\delta(\cdot)$ is also added to the model. Let $\eta(\cdot)$ 
represent a Gaussian random effect, define $\bu = \{\bu_1,\ldots,\bu_m\}$ 
to be the knots, and let $S_k(\cdot)$ define the basis function corresponding 
to the $k$th knot. Under the SME model, the usual spatial process $f(\cdot)$ 
is then replaced by the linear combination of basis functions and the fine-scale 
variation process. With a linear mean structure, the model for a single 
observation $\bs_i$ is 
\begin{equation}
\label{sremodel}
	y(\bs_i) = \bx(\bs_i)'\bbeta + \sum_{k=1}^{m}S_k(\bs_i)\eta(\bu_k) +
	 \delta(\bs_i) + \epsilon(\bs_i),
\end{equation}
where $\bx(\cdot)=[x_1(\cdot),\ldots,x_p(\cdot)]'$ is a vector of known covariates 
with (unknown) coefficients $\bbeta=(\beta_1,\ldots,\beta_p)'$. 
Also, let $\bseta(\bu)\sim N(\bzero,\bK)$, $\bdelta(\bs) \sim N(\bzero, \sigma_{\delta}^2\bV_{\delta})$, 
and $\bepsilon(\bs) \sim N(\bzero,\sigma_{\epsilon}^2\bV_{\epsilon})$
be mutually independent 
with $\bV_{\delta}$ and $\bV_{\epsilon}$ known diagonal matrices with
entries corresponding to the fine-scale and measurement error
variances, respectively. These diagonal elements are realizations of
known functions $v_{\delta}(\cdot)$ and $v_{\epsilon}(\cdot)$. 

Define the dispersion matrix of $y(\cdot)$ by $\bSigma =
\bS\bK\bS'+\sigma_{\delta}^2\bV_{\delta}+\sigma_{\epsilon}^2\bV_{\epsilon}$ with $\bS$ 
denoting the full $n\times m$-matrix of basis functions with $k$th column given by
$\bS_k(\bs)$. Defining $\bX$ as the design matrix for the mean structure,  
the log-likelihood for $y(\cdot)$ in simplified matrix notation is then the usual
\begin{eqnarray}
\label{loglike}
	\ell(\bK,\sigma_{\delta}^2,\sigma_{\epsilon}^2,\bbeta;\by) &\propto& -\frac{1}{2}
	(\by - \bX\bbeta)'\bSigma^{-1}(\by - \bX\bbeta) \nonumber \\
	& & - \frac{1}{2}\log(|\bSigma|).
\end{eqnarray}
Spatial predictions require the covariance between the set of 
observations and the desired locations, $\mbox{Cov}[\by(\bs_0), \by(\bs)]$. 
Let $\bA = [\bS_1(\bs_0),\ldots,\bS_m(\bs_0)]'$ be the $N \times m$ matrix of basis functions 
relating the knots to the desired locations and define $\bI_s$ as an $N \times n$ matrix with 
entries equal to $I_s(i,j) = I[\bs_{0i} = \bs_j]$. We then define the covariance,
\begin{equation}
\label{Cs0}
	\bC(\bs_0) = \mbox{Cov}[\by(\bs_0), \by(\bs)] = \bA\bK\bS' + 
	\sigma_{\delta}^2v_{\delta}(\bs_0)\bI_s
\end{equation} 
and the kriging estimates -- see Section 3.4.5 of 
\cite{cressie1993} -- as
\begin{equation}
\label{frk}
 \hat\by(\bs_0) = \bX(\bs_0)\hat\bbeta + \bC(\bs_0) \bSigma^{-1}(\by - \bX\hat\bbeta)
\end{equation}
where $\hat\bbeta = (\bX'\bSigma^{-1}\bX)^{-1}\bX'\bSigma^{-1}\by$.
The kriging standard error (KSE) is given by
$\hat\bsigma_k(\bs_0) = \{\bA\bK\bA' + \sigma_{\delta}^2\bV_{\delta}(\bs_0) 
- \bC(\bs_0)\bSigma^{-1}\bC(\bs_0)' 
+ [\bX(\bs_0) - \bX'\bSigma^{-1}\bC(\bs_0)']' \linebreak
(\bX'\bSigma^{-1}\bX)^{-1} [\bX(\bs_0) - \bX'\bSigma^{-1}\bC(\bs_0)'] \}^{\frac{1}{2}}$ 
\cite{cressiejohannesson2008} where $\bV_{\delta}(\bs_0)$ is diagonal matrix 
with diagonal entries given by the function $v_{\delta}(\cdot)$
evaluated at $\bs_0$.  
By defining $\bD = \sigma_{\delta}^2\bV_{\delta}+\sigma_{\epsilon}^2\bV_{\epsilon}$, 
the computational burden of matrix inversion is reduced using the identity
\begin{equation}
\label{SMW}
	\bSigma^{-1} = \bD^{-1} - \bD^{-1}\bS[\bK^{-1} + \bS'\bD^{-1}\bS]^{-1}S'\bD^{-1},
\end{equation}
which follows from the Sherman-Morrison-Woodbury formula
\cite{hendersonsearle1981}. Note that the inversion of 
the $n \times n$ matrix $\bSigma^{-1}$ is replaced by that of two
(smaller) $m \times m$ matrices and the diagonal matrix, $\bD$. \cite{cressiejohannesson2008}
recommended the Method of Moments estimator for the parameters
in order to guarantee positive definiteness of $\hat\bK$, the 
estimate of $\bK$.

An alternative approach \cite{katzfusscressie2011} obtained maximum
likelihood estimates (MLEs) using  the 
Expectation-Maximization (EM) algorithm \cite{dempsteretal77} with
$\by$ treated as the observed data and  $\bseta$ and $\bepsilon$ as 
the missing data. In the absence of independent information regarding 
the measurement error of the process, \cite{kangcressieshi2010} provide 
a method of estimating $\sigma_{\epsilon}^2$ that involves extrapolating 
the semivariogram back to the origin. Therefore, \cite{katzfusscressie2011}
assume that $\sigma_{\epsilon}^2$ is known or can be obtained
independently. We adopt the same approach in this paper. Starting from
initial values, the variance parameter estimates are obtained upon updating 
\begin{eqnarray}
\label{M}
  \bK^{(t+1)} &=& \bK^{(t)} - \bK^{(t)}\bS'{\bSigma^{(t)}}^{-1}\bS\bK^{(t)} \nonumber \\ 
  &+& [\bK^{(t)}\bS'{\bSigma^{(t)}}^{-1}(\by - \bX\hat\bbeta)] \nonumber \\
  & & [\bK^{(t)}\bS'{\bSigma^{(t)}}^{-1}(\by - \bX\hat\bbeta)]' \nonumber \\
	\sigma_{\delta}^{2(t+1)} &=& \sigma_{\delta}^{2(t)} + \frac{(\sigma_{\delta}^{2(t)})^2}{n}\tr[{\bSigma^{(t)}}^{-1}((\by - \bX\hat\bbeta) \nonumber \\
	& & (\by - \bX\hat\bbeta)'{\bSigma^{(t)}}^{-1} - \bI) \bV_{\delta}], \nonumber \\ 
\end{eqnarray}
from the $t$th to the $(t+1)$th iteration, and proceeding till
convergence. The estimate of $\bbeta$ denoted by $\hat{\bbeta}$  is  
obtained using generalized least squares (GLS) as in (\ref{frk}).

This methodology assumes that the basis functions are fully specified smoothing 
functions. \cite{cressiejohannesson2008} and \cite{katzfusscressie2011} use 
the local bisquare function. We provide further details in
Section \ref{basis} but note that this 
function requires a tuning parameter, which the authors set at a
constant 1.5 times  
the minimum distance between the inter-knot distance. In the context
of spatial statistics, this tuning  
parameter has interpretation similar to the range parameter and so estimation 
provides insight into the rate of decay of spatial dependence.
Further, inaccurate specification of this parameter has the potential
to  also result in poor  prediction, therefore our methodology details
an extension to the EM algorithm  
that allows for estimation of a ``range'' parameter within $\bS$.

This paper is organized as follows. In Section 2 of this paper 
we propose a combination of concepts from the EM approach to fixed
rank kriging (FRK) and the Alternating Expectation Conditional Maximization algorithm (AECM) 
\cite{mengvandyke1997,chenmaitra2011} that allows for 
computationally practical estimation of a 
continuous parameter within the basis functions, 
similar to the use of the ECM for spatio-temporal problems 
by \cite{xuwikle2007}. Results of extensive 
simulation-based evaluations of our methodology in
Section 3 show that estimation of the basis functions 
can improve prediction and is robust against certain model misspecification. 
Section 4 demonstrates the applicability of our 
methodology to predicting temperatures across the US. We conclude with 
some discussion and pointers to future work.


\section{Methodology}
\label{methodology}

\subsection{Choice of Basis Functions}
\label{basis}
Fixed rank kriging and the proceeding EM estimation methodology 
recommend the use of multi-resolution basis functions in order to 
capture different scales of spatial variation. In particular, they 
use the local bisquare function which defines the 
basis function at the $l$th resolution as 
\begin{equation}
\label{B}
  S_{k(l)}(\bx) = \Psi \left( \frac{\|\bx - \bu_{k(l)}\|}{r_l} \right)
  \quad \forall\mbox{ }\bx \in D,
\end{equation}
where $\bu_{k(l)}$ is a knot assigned to the $l$th resolution, with 
$l = \{1,\ldots,L\}$. $\Psi(d)$ is the local bisquare function defined as
\begin{equation}
  \label{bisquare}
	\Psi(d) = \left\{ 
	\begin{array}{l l}
		\left(1 - d^2 \right)^2  & \quad 0 \le d \le 1 \\
		0 & \quad d > 1 \\
	\end{array} \right. 
\end{equation}
where $r_l = b \hspace{1 mm} \mbox{min}\{\|\bu_{i(l)}-\bu_{j(l)}\| : j \neq i, 1 \le i,j \le m\}$ 
and $b$ is some constant. We follow the recommendation and notation of these 
papers by defining our matrix of basis functions $\bS$ by \eqref{B} and \eqref{bisquare}. 
This form is particularly useful in that (\ref{bisquare}) sets any value 
equal to zero where the distance between the location and the knot is 
greater than $r_l$, the ``bandwidth''. This allows the opportunity to utilize 
matrix operations and algorithms (see {\em e.g.} \cite{davis2006}) that have been specifically 
designed to exploit sparsity. The distances are often defined in terms of the 
Euclidean norm, however this is not a necessary condition, as shown in our 
application to temperature data.

In practice, the number of knots used in prediction will be a function of the 
computational resources available and, consequently, may be considered
to be known. 
From (\ref{B}), it is clear that the remaining unknowns are the resolution $l$
and the bandwidth constant $b$. The optimal resolution is one of a finite, and 
generally small, set of positive integers. Given a finite set, estimation and 
prediction with varying resolutions remain easily parallelizable processes
and so a model selection approach comparing performance of varying levels of 
resolution can be implemented. The domain of $b$ is $\mbox{\R}^+$, so a model 
selection approach would require a discretized domain. This approach
can provide an 
estimate for $b$ that will produce reasonable predictions, however numerous 
estimation chains may be necessary and if the set of \emph{a priori} selected 
possible values for $b$ does not encompass the true value, the estimate will 
be biased. Maintaining a model selection approach to determining the optimal 
resolution, $l$, we now direct our focus to improved estimation of $b$.

\subsection{Restricted Maximum Likelihood Estimation of Bandwidth}
\label{mleb}
This section develops methodology for obtaining a restricted maximum likelihood (REML) 
estimate of $b$ along with those for $\bK$ and $\sigma_{\delta}^2$. As noted by 
\cite{katzfusscressie2009}, an analytic solution to ML 
estimation of the parameters in the SME model does not appear to exist 
and direct numerical ML estimation is also challenging in that it 
requires maintaining a positive definite covariance matrix $\bK$. For 
these reasons, the EM algorithm was used for parameter estimation.

When implementing the EM algorithm to estimate $b$, we wish to keep 
the advantageous structure of $\bS$ intact. To avoid singularity issues 
during estimation, we also assume that $\sigma_\epsilon^2$ is
known and that GLS is used to estimate $\bbeta$, as before. Thus, we follow 
\cite{katzfusscressie2011} and the remaining two random variables 
play the role of ``missing values'': $\bseta$ and $\bdelta$. Assuming
independence of $\bseta \sim N(\bzero, \bK)$ and $\bdelta \sim
N(\bzero, \sigma_{\delta}^2\bV_{\delta})$, we denote  
$\btheta^{[t]}=\{\bK^{[t]}, \sigma_{\delta}^{2[t]}, b^{[t]}\}$ as the parameter
values at the $t$th iteration. The analytical solution to the
M-step of the EM algorithm then requires maximizing  
\begin{equation}
\label{EMlike}
\begin{split}
Q(\btheta; \btheta^{[t]})  = & -\frac{1}{2}\{\log|\bK| + \tr(\bK^{-1} 
	E_{\btheta^{[t]}}[\bseta\bseta'\mid\by]) \\
	& + \log(\sigma_{\delta}^{2n}|\bV_{\delta}|) + \frac{1}{\sigma_{\delta}^2}\tr(\bV_{\delta}^{-1} E_{\btheta^{[t]}}[\bdelta\bdelta'\mid\by]) \\
    & + \frac{1}{\sigma_{\epsilon}^2} \tr(\bV_{\epsilon}^{-1} 
	[\by - \bX\bbeta - \bS E_{\btheta^{[t]}}[\bseta\mid\by] \\ 
	& - E_{\btheta^{[t]}}[\bdelta\mid\by]]  [\by - \bX\bbeta - \bS E_{\btheta^{[t]}}[\bseta\mid\by] \\
	& - E_{\btheta^{[t]}}[\bdelta\mid\by]]') \}
\end{split}
\end{equation}

Maximizing \eqref{EMlike} involves taking partial derivatives with
respect to each parameter. For $\bK$ and $\sigma_{\delta}^2$, this is fairly 
straightforward as these parameters appear separately within the summation. 
Thus, the resulting updating scheme includes the equations found in \eqref{M}. 
Partial differentiation with respect to $b$, however, involves 
taking the derivative of a quartic term nested within the matrix of 
multi-resolution basis functions. An analytic solution does not exist,
and the usual EM algorithm does not appear to show much promise. So
we develop an AECM algorithm to estimate $b$. 

\subsection{Estimating Bandwidth through AECM}
\label{aecm}
In order to exploit the analytical iterative updating scheme for $\bK$ and 
$\sigma_{\delta}^2$, we partition the parameters as $\btheta_1 = \{\bK, 
\sigma_{\delta}^2 \}$ and $\theta_2 = \{b\}$. The CM-step of the AECM 
algorithm alternates between maximizing the likelihood function (\ref{EMlike}) 
with respect to $\btheta_1$ (keeping $\theta_2$ fixed at its current value) 
and $\theta_2$ (by holding $\btheta_1$ fixed at its current value). $\btheta_1$ 
is, therefore, updated by (\ref{M}) as before. An estimate for
$\theta_2$, however,  
cannot be calculated analytically nor numerically, so in lieu of maximizing 
(\ref{EMlike}), we maximize the restricted log-likelihood function given by 
\begin{equation}
\label{remllogl}
\begin{split}
	\ell(\bK,\sigma_{\delta}^2,b; \by,\sigma_{\epsilon}^2) \propto &  -\frac{1}{2}
	(\by - \bX\hat{\bbeta})'\bSigma^{-1}(\by - \bX\hat{\bbeta}) \\
	& - \frac{1}{2}\log(|\bSigma|) - \frac{1}{2}\log(|\bX'\bSigma^{-1}\bX|). 
\end{split}
\end{equation}

When restricted to be a function of $\theta_2$, (\ref{remllogl}) involves 
a covariance matrix which is in a form that allows for the 
use of dimension reduction equalities in the matrix calculations. As noted in 
\cite{cressiejohannesson2008}, the Sherman-Morrison-Woodbury formula can be 
used to obtain a computationally efficient form of $\bSigma^{-1}$. 
To avoid the burdensome computation of the determinant, we also 
invoke  Sylvester's 
determinant theorem \cite{harville2008,cressiejohannesson2008}, which is an analogue to the 
Sherman-Morrison-Woodbury formula. Let $\bX$ be an invertible $n \times n$ matrix, 
$\bA$ an $n \times m$ matrix, $\bB$ an $m \times n$ matrix, and $\bI_m$ is the 
$m \times m$ identity matrix then   
$\det(\bX + \bA\bB) = \det(\bX) \det(\bI_m + \bB\bX^{-1}\bA)$. Let $\bC$ be the 
Cholesky decomposition of $\bK$ ($\bK=\bC\bC'$). This theorem along
with the fact that  $\bK$ is positive definite and some algebra takes us
to the reduction
\begin{equation}
\label{slydet}
	\det(\bD + \bS\bK\bS') = \det(\bD) \det(\bC)^2 \det(\bK^{-1} + \bS'\bD^{-1}\bS)
\end{equation} 
which provides us with the advantage that it requires computing an 
inverse of an $m \times m$ matrix and a determinant of an $m \times m$ matrix 
instead of the determinant of an $n \times n$ matrix. Also, a Cholesky 
decomposition of the two $m \times m$ matrices, $\bK$ and $\bK^{-1} + 
\bS'\bD^{-1}\bS$, can be used in both the calculation of
the determinant using \eqref{slydet} and the inverse covariance in the
likelihood that uses \eqref{SMW}, providing an additional reduction in
computation. Finally, since $\bD$ is a  
diagonal matrix and $\bC$ is a lower triangular matrix, the determinant of 
each is simply (and efficiently) computed as the product of the diagonal
elements. Let $\bC_2$ be the Cholesky decomposition of $(\bK^{-1} +
\bS'\bD^{-1}\bS)$ and $\bC_3$ be the Cholesky decomposition of $\bX'\bSigma^{-1}\bX$. 
The final form of the restricted log-likelihood is then
\begin{equation}
\label{altlike}
\begin{split}
	\ell(\bK,\sigma_{\delta}^2,b; \by,\sigma_{\epsilon}^2) \propto& 
	- \frac12(\by - \bX\hat{\bbeta})'\bSigma^{-1}(\by - \bX\hat{\bbeta}) \\
	&- \frac 12\log(|\bD|) - \log(|\bC|) \\
	&- \log(|\bC_2|) - \log(|\bC_3|)
\end{split}
\end{equation}
where $\bSigma^{-1}$ is defined as in (\ref{SMW}). Parameter estimates
using our AECM algorithm are obtained by computing $\btheta_1$ from \eqref{M}
given $\theta_2^{[t]}$, then maximizing (\ref{altlike}) with respect to 
$\theta_2$ given $\btheta_1^{[t]}$, and repeating this process until convergence.

This maximization step requires solving at  most an $m \times m$
linear system of equations; however, to perform a full maximization 
of $\theta_2$ between each update of $\btheta_1^{[t]}$ would be 
immensely inefficient. Numerical maximization techniques for $\theta_2$ 
may also fail when the log-likelihood function is ill-behaved. 
However, exploratory results suggest that the function remains locally 
quadratic in $b$ near it's MLE, so we use a search algorithm 
to explore the log-likelihood for its maximum with respect to $b$.

Given a quadratic function and a desired maximum, we can use the Quadratic 
Search algorithm \cite{muuquy2003} that requires only three evaluations of 
the likelihood. In order to improve initial starting values, we propose 
using the Golden Search algorithm \cite{kiefer1953, avrielwilde1966} to 
provide a ``burn-in'' phase. Also, since the overall discrepancy between 
the starting values and the final estimates of $\bK$ could be very large, 
it is best to obtain a reasonable estimate of $\btheta_1$ before any 
updating of $\theta_2$. So we also suggest a burn-in phase of the EM 
algorithm at a few values of $b$ to begin. The full updating algorithm 
is therefore as follows:

\begin{enumerate}
	\item Obtain $\btheta_1$ by \eqref{M} until weak convergence.
	\item \begin{enumerate}
				\item Set $\btheta_2^{[t]}=\{b_1,b_2,b_3,b_4\}$ based on the Golden Search algorithm.
				\item Set $\btheta_2^{[t]}=\{b_1,b_2,b_3\}$ based on the Quadratic Search algorithm.
				\end{enumerate}
	\item Obtain $\btheta_1$ by \eqref{M} for each $\btheta_2^{[t]}$.
	\item Evaluate (\ref{altlike}) for each set of $\btheta_2^{[t]}$ and $\btheta_1^{[t]}$.
	\item Set $\btheta^{[t+1]}$ to the set of parameter values that maximize \eqref{altlike}.
	\item Repeat 2(a)-5 until weak convergence.
	\item Repeat 2(b)-5 until convergence.
\end{enumerate}

The convergence criteria for the initial burn-in phase of the EM 
algorithm need not be very strict, as the likelihood function 
typically plateaus after only a few iterations. However, obtaining 
better initial values for $\btheta_1$ is still advisable to avoid the
problem of our log-likelihood getting trapped in a local maxima. Also,
\cite{cressiejohannesson2008}  
suggest using $b=1.5$ in the bisquare basis functions, so we will use 
a set of initial values for $b$ that encompasses 1.5 for the EM algorithm 
burn-in phase.
In the next section, we investigate, through simulation, the impact 
of estimating the bandwidth constant, $b$, of the local bisquare basis 
functions.


\section{Experimental Evaluations}
\label{experiments}

\subsection{Experimental Setup}
\label{exp_setup}
Section 2 outlined an AECM algorithm for efficiently
computing an MLE for $b$ in addition to that for $\bK$ and
$\sigma_{\delta}^2$. In this section, we evaluate performance of our
method through a 
series of simulation experiments using the local bisquare basis
function of \eqref{B}.  Our experimental design was similar to that
used by \cite{katzfusscressie2011},  
which followed \eqref{sremodel} as the generative model. For our
experiments, we chose a 
one-dimensional spatial domain,  
$\mathcal{D} = \{1,\ldots,256\}$, with $n=64$ observed locations
selected either completely randomly or following a ``clustered''
random sample. The clusters  
were created by splitting $\mathcal{D}$ into intervals of 32 locations and 
omitting every other interval. To elucidate, a clustered random sample is a 
subset of $\{1,\ldots,32\} \cap \{64,\ldots,96\} \cap \{128,\ldots,160\} \cap 
\{192,\ldots,224\}$. Five knots at a single resolution were used, 
centered at $\{0.5, 64.5, 128.5, 192.5, 256.5\}$. 

The main focus of our simulation was to understand the effects on
estimation and 
prediction when aspects of the covariance structure are varied. To that end, 
we used a simple linear mean structure ($\beta_0 = 5$ and
$\beta_1=0.08$) for all simulations. To cover the wide variety of 
covariance matrices $\bK$ allowed by FRK, we simulated covariance
matrices to have stationary, positively-correlated nonstationary, and
unrestricted nonstationary structures. The stationary covariance was
modeled after a Mat\'ern covariance  
matrix \cite{stein1999} which, for locations $\bs_i$ and $\bs_j$, is defined by
\begin{equation}
\label{matern}
\begin{split}
    Cov[\bff(\bs_i), \bff(\bs_j)] =&	\frac{\rho}{2^{\nu - 1} \Gamma(\nu)} \left(\frac{\|\bs_i - \bs_j\|}{\theta}\right)^{\nu} \\
	& K_{\nu}\left(\frac{\|\bs_i - \bs_j\|}{\theta}\right), 
\end{split}
\end{equation}
where $K_{\nu}$ is the modified Bessel function of the second kind of
order $\nu$, $\nu$ is the smoothness parameter, $\rho$ is the sill 
or scale parameter, and $\theta$ is the range parameter, with 
$\nu,\theta \in (0,\infty)$ and $\rho \in [0,\infty)$. For our
  simulation setup, we 
held the parameters fixed at $\rho = 9$, $\theta = 96$, and $\nu = 1$.
The unrestricted nonstationary structure was created through the matrix product of a 
realization from the \cite{wishart1928} distribution ($\bW \sim W_5(2\bI_5, 
10)$, where $\bI_5$ is the $5 \times 5$ identity matrix) sandwiched in between diagonal 
matrices with elements increasing from 1 to 5. This simulation procedure created 
significant nonstationarity with expected value given by a diagonal
matrix with elements  
$\{2,8,18,32,50\}$. The positively correlated nonstationary structure
was simply the absolute value of the unrestricted nonstationary matrix
described above. 

Modeling the covariance as spatial in nature is only useful when
significant spatial  
dependence exists in the data. To quantify the magnitude of spatial
structure, we compare  
two ratios of variances: (i) $\tr (\bS\bK\bS' + \sigma_{\delta}^2\bV_{\delta}) / 
\tr(\sigma_{\epsilon}^2\bV_{\epsilon})$, the signal-to-noise ratio, and (ii) 
$\tr (\sigma_{\delta}^2\bV_{\delta}) / \tr (\bS\bK\bS' + \sigma_{\delta}^2\bV_{\delta})$, 
the fine-scale variation proportion. Testing the limits of our methodology, 
we varied the fine-scale variation parameter  
$\sigma_{\delta}^2$ to be within $\{0.01, 0.1, 1\}$ and the
measurement error   
$\sigma_{\epsilon}^2$ to lie within $\{1, 10, 100\}$. This resulted in
the ratio (i) to range between $\{0.037, 30.96\}$ and for ratio (ii)
to range 
between $\{0.0003, 0.215\}$. 
Additional simulation inputs included the bandwidth constant ($b =
\{0.5, 1, 1.5, 2\}$),  
the number of resolutions ($l = 1$), and the weighting functions 
($v_{\delta}(\cdot) = v_{\epsilon}(\cdot) \equiv 1$). 

The \texttt{R} \cite{cran2015} package {\tt fields}
\cite{fields2015} was used  
to simulate 1000 Gaussian random fields for each combination of
covariance type,  
parameter value, and sampling design. For each simulated field, we obtained 
MLEs using both our AECM algorithm (estimating $b$) and the EM algorithm 
(leaving $b=1.5$ fixed). From these estimates, we compared model fits
quantified by \cite{kullbackleibler1951}'s K-L divergence and 
calculated the corresponding kriging estimates and standard errors at all 
locations within $\mD$.

\subsection{Results}
Computational cost is crucial for any spatial prediction method 
intended for massive datasets. The simulation results, however, were generated 
using a relatively small sample size in order to permit evaluation of
the methodology in a wide range of settings. Without implementing parallel 
processing techniques, the computational complexity of our method is a known 
$O(3nr^2)$ versus $O(nr^2)$ for the EM 
approach, after burn-in. A thorough comparison of computational cost is, 
therefore, saved for the application in Section 4 where
the sample size is larger.  

The other relevant points of comparison in our simulation experiments
are parameter estimation error, model fit, and performance in
prediction.  We assess  error in parameter estimation by calculating
the median absolute deviation (MAD) given the true parameter values. 

\begin{figure}[htbp]
  \mbox{
    \subfloat[$\hat\beta_0$]{\includegraphics[width=0.5\columnwidth]{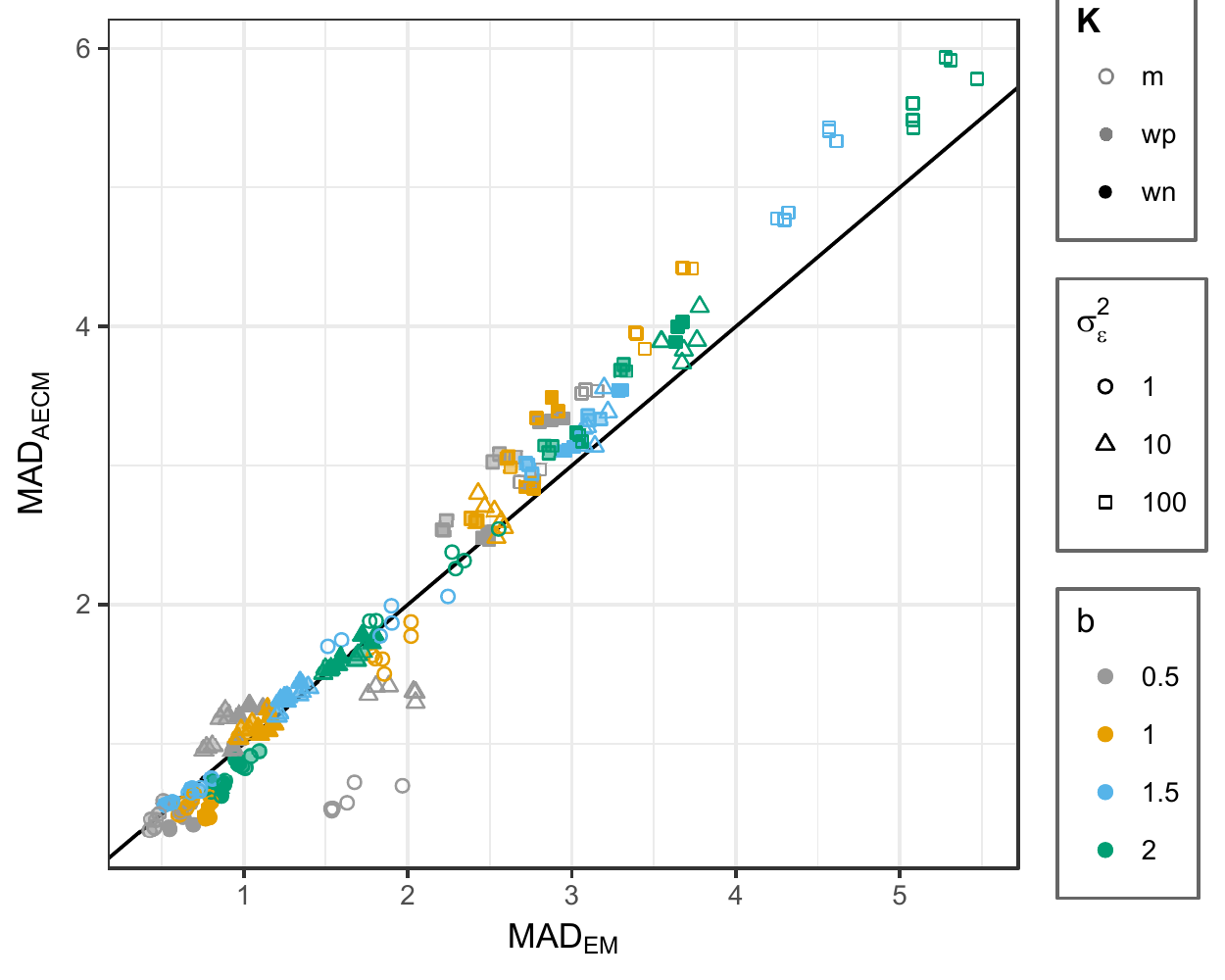}}
    \subfloat[$\hat\beta_1$]{\includegraphics[width=0.5\columnwidth]{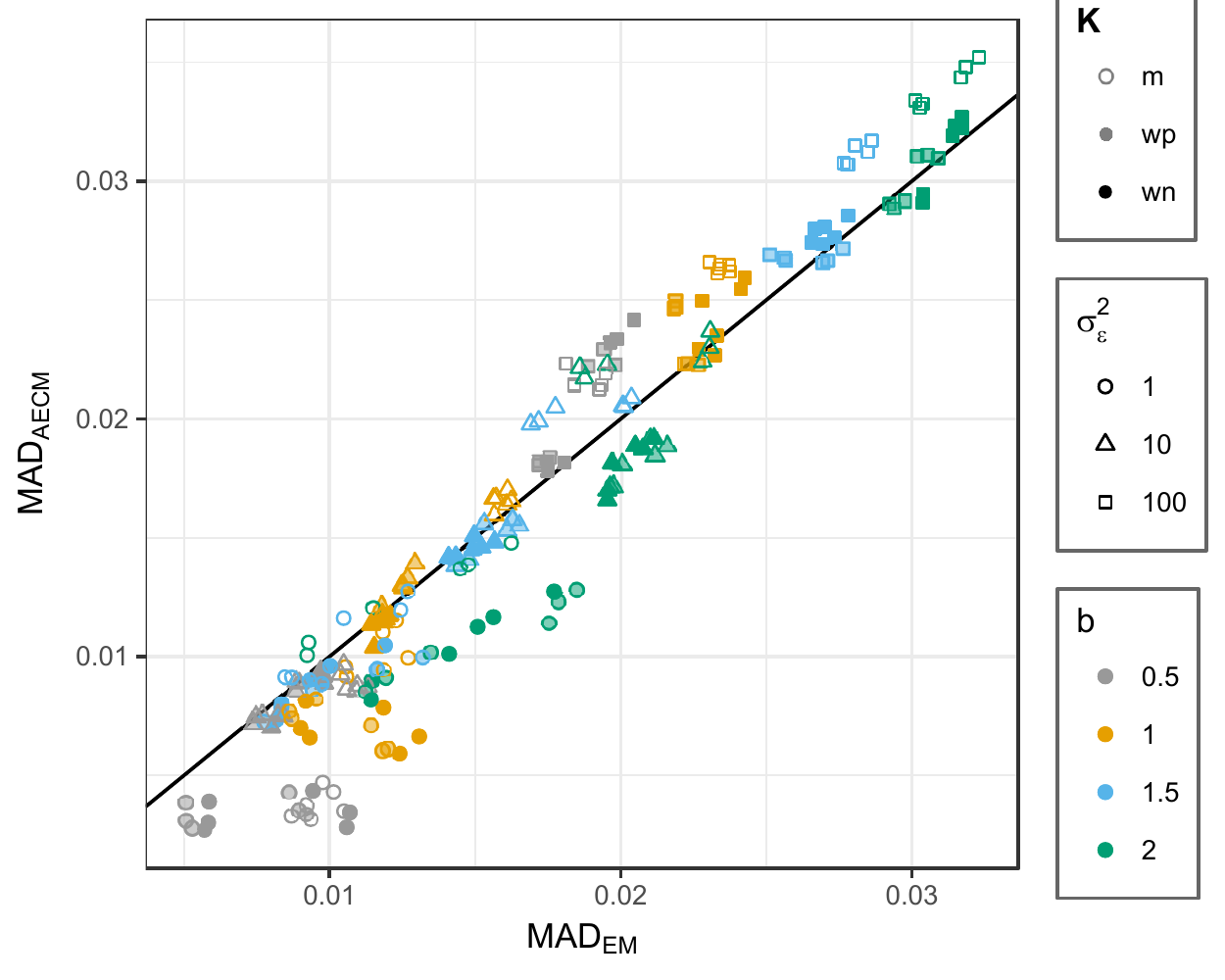}}
  }
  \mbox{\subfloat[$\hat\bK$]{\label{Kplot} \includegraphics[width=0.5\columnwidth]{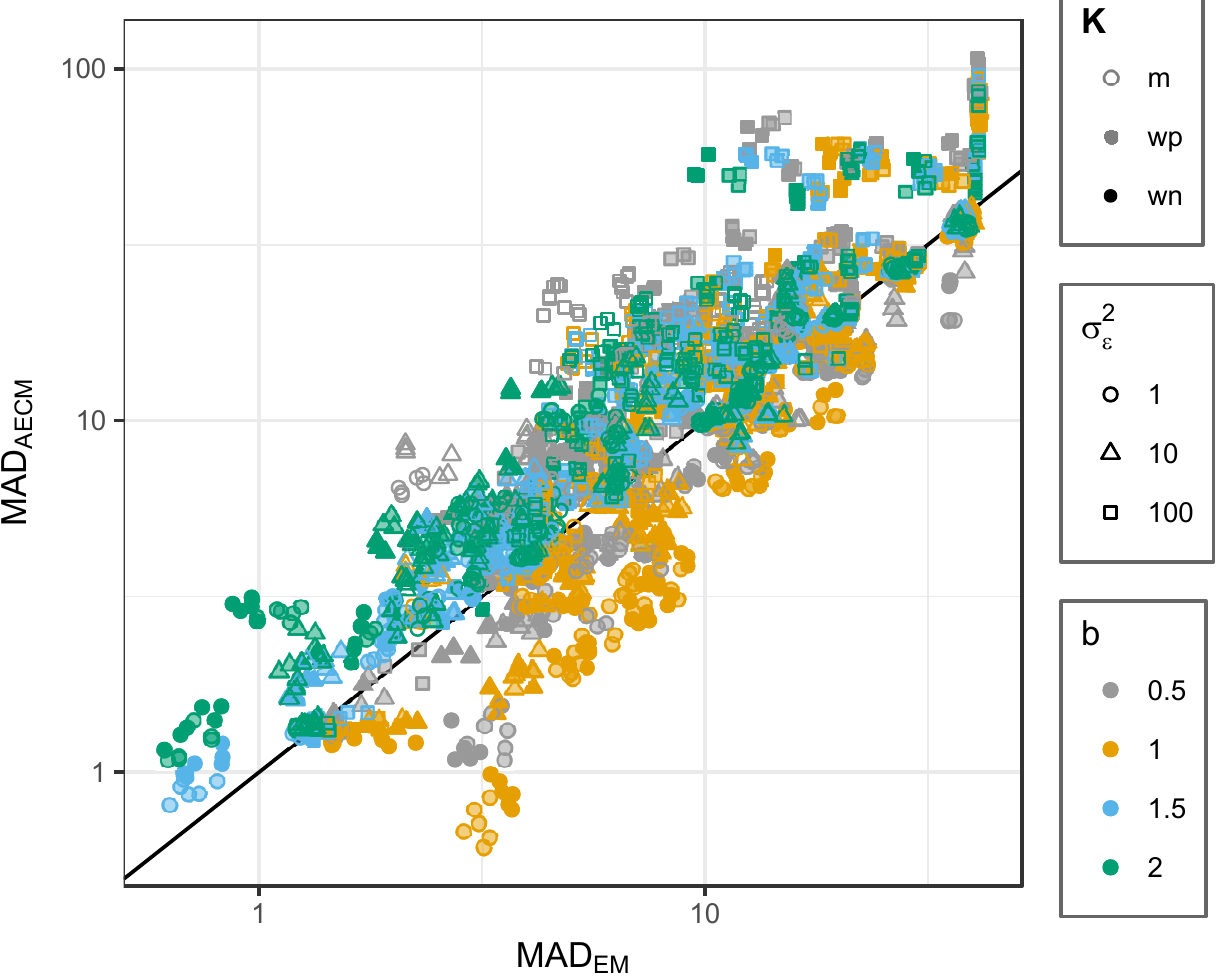}}
    \subfloat[$\hat\sigma_{\delta}^2$]{\includegraphics[width=0.5\columnwidth]{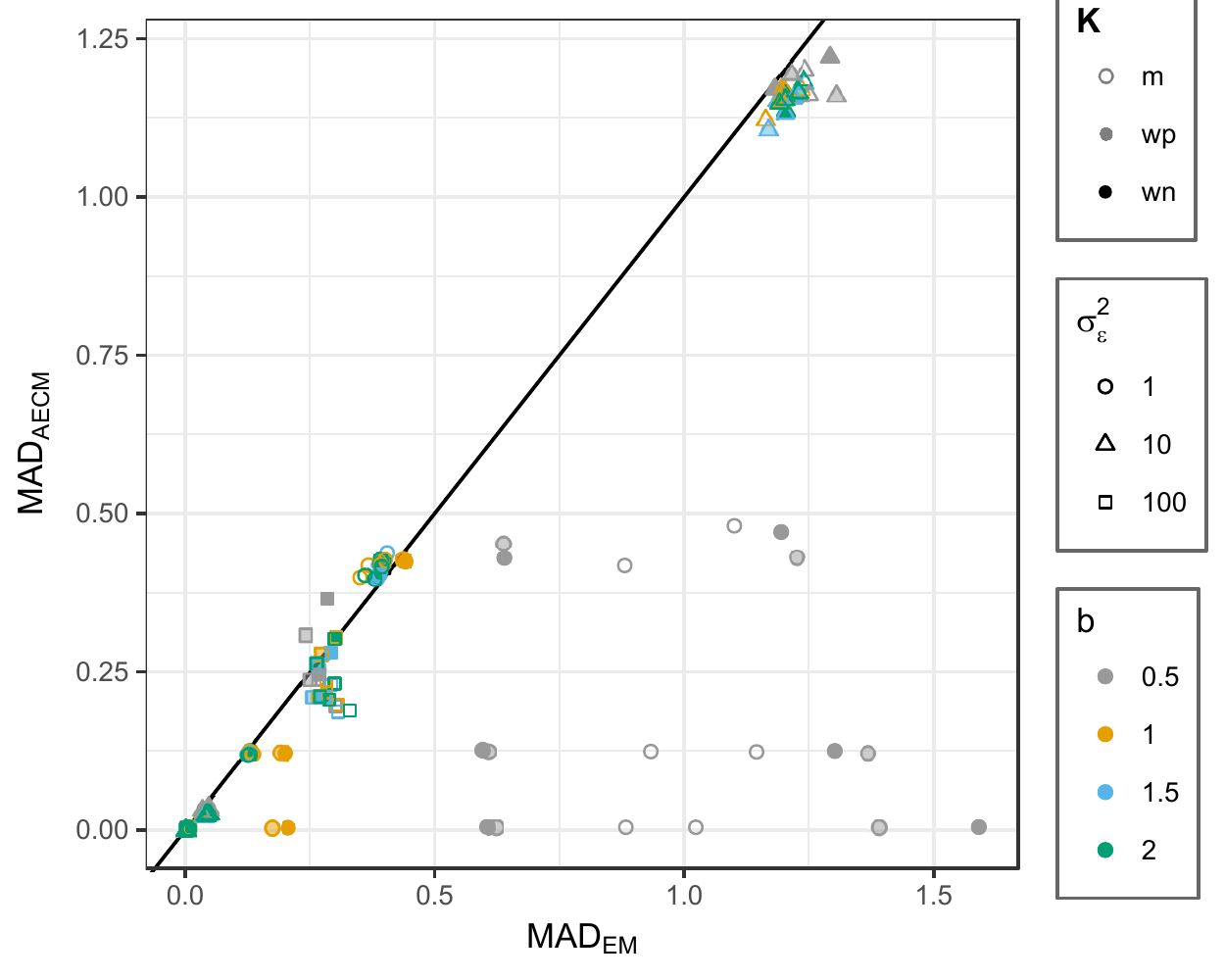}}}
  \caption{Median absolute deviation of parameter estimates under the 
    various combinations of input as described in \ref{exp_setup}. The 
    covariance types for $\bK$ are abbreviated m = Mat\'ern, wp = Wishart 
    positive, and wn = Wishart negative.}
  \label{parsplot}
\end{figure}

Figure \ref{parsplot} displays the results: we see that parameters
that are part of the mean structure exhibit similar patterns.  
The error in slope and intercept estimation is decreased by the AECM
algorithm in  
the presence of low measurement variance ($\sigma_{\epsilon}^2 = 1$) and 
increased given high measurement variance ($\sigma_{\epsilon}^2 = 100$). The 
magnitude of deviation is also an increasing function of measurement error for 
either approach. The AECM algorithm decreases the error in estimating $\beta_0$ 
when the true value of $b$ is not equal to 1.5, provided the
measurement error is  
low. When $\sigma_{\epsilon}^2 = 10$, both approaches perform similarly and when 
$\sigma_{\epsilon}^2 = 100$, again, the AECM algorithm exhibits signs of 
overfitting. This pattern is also present in estimation of $\beta_1$,
however the  relative improvement by the AECM algorithm is larger. When 
$\sigma_{\epsilon}^2$ is less than 100 and the covariance of $\bK$ is 
nonstationary, the AECM algorithm consistently improves estimation of
$\beta_1$. These results hold even when we vary covariance type,
fine-scale variation, and sampling design.

Figure \ref{Kplot} shows the MAD for all unique elements of $\hat{\bK}$. 
Most are fairly similar between the AECM and EM approaches, but there
are some overall patterns that we notice. Specifically, in the
presence of large measurement error or when  
the true $b$ is at least 1.5, the EM algorithm reduces the MAD. When the true 
spatial range is small and measurement error does not overwhelm the signal 
({\em i.e.}, $b\le1$ and $\sigma_{\epsilon}^2\le10$) however, the AECM algorithm typically 
performs better. Surprisingly, this reduction in MAD is more consistent when 
$b=0.5$ than when $b=1$. Again, these trends hold irrespective of
covariance type, fine-scale variation, and  sampling design.

Accurate estimation of variance parameters in spatial fields can often be 
challenging. In particular, the fine-scale variation parameter in the
SME model is often underestimated. However, allowing a more flexible spatial  
structure by estimating the range parameter with respect to the knots
($b$) provides consistent reduction in the MAD for
$\sigma_{\delta}^2$.  
Only when both independent variance components are equal ($\sigma_{\delta}^2 = 
\sigma_{\epsilon}^2$) and occasionally when the measurement error is high does 
the AECM method fail to improve estimation.

\begin{figure}[htbp]
	\centering
	\includegraphics[width=0.8\columnwidth,angle=270]{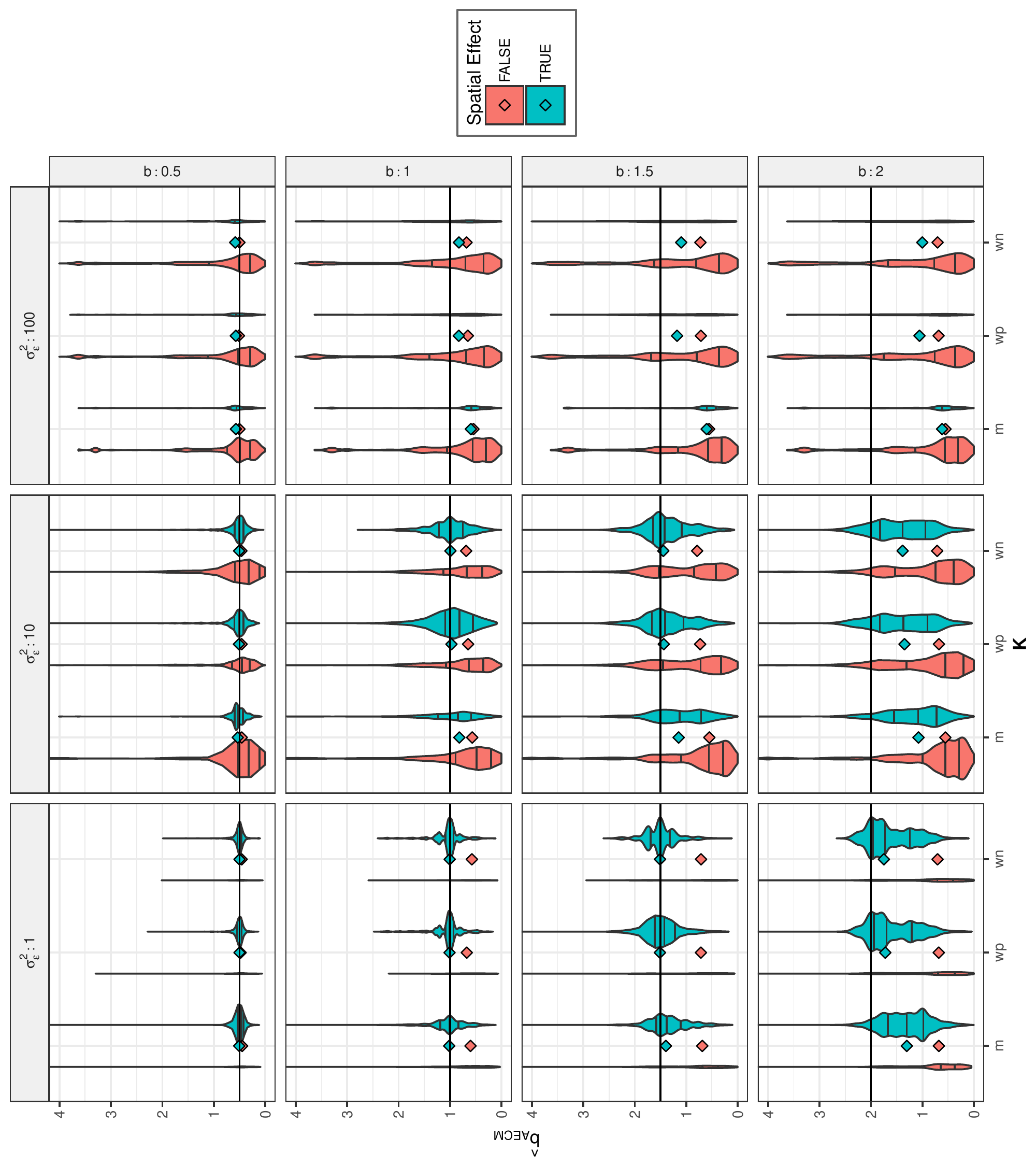}
	\caption{Distributions of $\hat{b}$ provided by the AECM. The mean of each 
		distribution is identified by a diamond and its quartiles by horizontal 
		black lines.}
	\label{bplot}
\end{figure}

Clearly, when observations contain extreme levels of measurement
error, parameter  
estimation given a SME model is problematic. Assuming any 
spatial model when non-spatial variance components dominate the total variance 
can have ill-effects on estimation accuracy. This is no more evident than when 
exploring distributions of $\hat{b}$ obtained through simulation, provided in 
Figure \ref{bplot}. Distributions are further segregated and color
coded between those that showed signification spatial autocorrelation
($\alpha < .05$) as measured  
by \cite{moran1950}'s I and those that did not. 
(Note that some distributions of $\hat{b}$ are truncated in order to highlight the 
range of highest density within each distribution, however 
these estimates account for less than 0.17\% of the total
simulations.) We see in the figure that there is a clear distinction
between distributions with and without significant spatial  
autocorrelation. When $\sigma_{\epsilon}^2 = 100$, nearly all
simulated fields do not  
exhibit enough spatial dependence to compensate for the large
measurement error. Without  
significant spatial dependence, $\hat b$'s tend to be negatively biased with a 
large percentage of the distribution residing between 0 and 1, made worse when 
$\bK$ has a Mat\'ern structure. This pattern persists for all 
levels of measurement error, however  
the percentage of fields with insignificant spatial dependence
decreases with decreasing
$\sigma_{\epsilon}^2$. Estimation of $b$ performs fairly
well for true values  
of $b \le 1.5$, however when $b=2$, there is again a tendency for
$\hat{b}$ to be  underestimated. 

\begin{figure}[htbp]
	\centering
	\includegraphics[width=0.8\columnwidth,angle=270]{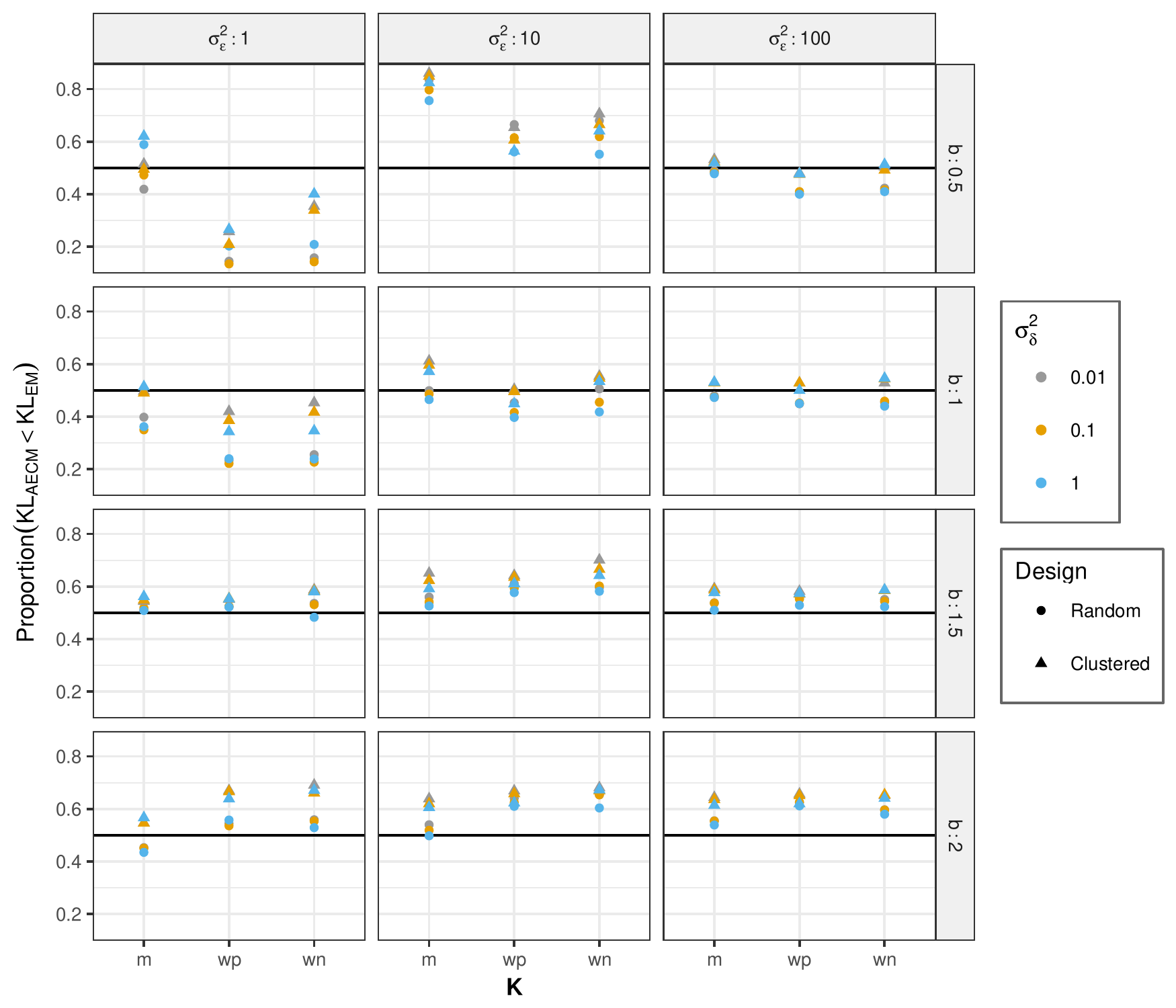}
	\caption{The proportion of simulations where the KL divergence was smaller for 
		the AECM method than the EM method.}
	\label{KLplot}
\end{figure}

The difference between the estimation methods clearly varies across parameters. 
To summarize the fit of all parameter estimates and, consequently, the overall 
model fit, we compared Kullback-Leibler (KL) divergence between the
two methods.  
KL divergence quantifies the amount of information lost when an estimated 
distribution, $Q$, is used in place of the true distribution, $P$.
The KL divergence~\cite{stein2014} for two multivariate normal distributions, 
$P \equiv N(\bmu_P, \bSigma_P) $ and $Q \equiv N(\bmu_Q, \bSigma_Q)$, on
$\mathbb{R}^n$ simplifies to   
$
	2\mbox{KL}(P,Q) = \tr(\bSigma_Q^{-1} \bSigma_P) + (\bmu_Q - \bmu_P)' \bSigma_Q^{-1}
	(\bmu_Q - \bmu_P) - \log \abs{\bSigma_P} + \log \abs{\bSigma_P} - n.
$
To summarize results, Figure \ref{KLplot} provides the proportion out of 1000 
simulations where the KL divergence is less for AECM than it is for
EM. Thus, values  over 0.5 indicate an improvement in overall model
fit when estimating $b$.

Interestingly, although individual parameter estimates  almost
unanimously improved using AECM when $b \le 1$ and
$\sigma_{\epsilon}^2 = 1$, the opposite holds when comparing KL  
divergence. In these cases, overall model fit is better assuming a 
fixed $b=1.5$. When $b \ge 1.5$, on the contrary, nearly every
combination of simulation  input resulted in a reduced KL divergence
using the AECM algorithm. In the presence of high measurement errors,
results were similar between estimating $b$ and leaving it fixed at
1.5, regardless of the true value of $b$. When $\sigma_{\epsilon}^2 =
10$, the AECM algorithm decreases KL divergence the most consistently,
reaching a proportion as high as 0.86.

Our final performance comparisons were in the context of prediction.  
To that end, we summarized our results into three metrics: mean square
prediction error (MSPE), ratio of KSE, and prediction interval coverage (PIC). 
MSPE is the usual statistical summary which calculates the average squared loss 
between the kriging predictions and the true value at unobserved locations. 
Since this is a simulation setting, we can define the true value as the uncorrupted 
spatial process, $\{ \by(\bs_0) - \bepsilon(\bs_0) \}$. 
The median KSE at each spatial location is computed given estimated parameters by either 
method. The ratio KSE (rKSE) is the ratio between this median KSE and
the KSE computed given the true parameter values at similar simulation
input values. Ideally, the rKSE is equal to 1. 
PIC is the percentage of prediction intervals that contain the true value, averaged over 
all locations. Since we assumed 95\% prediction intervals, we expect the PIC to be roughly 
0.95 for each case. For each summary statistic, the median was computed over all values of 
$\sigma_{\delta}^2$ and sampling designs and results are provided in
Table 1. 
Medians for every unique combination of input values are provided in
Section S-1.1 of the Supplementary Materials. 

\begin{table}
  \caption{\label{pred_sum} Median predictions results, comparing mean
    square prediction error (MSPE),
    ratio of median estimated versus true KSE, and prediction interval
    coverage (PIC).}
  \begin{tabular}{lll|rrr|rr|rrr}\hline 
    & & & & MSPE & & rKSE & & & PIC & \\
      K & $\sigma_{\epsilon}^2$ & $b$ & True & AE & EM & AE & EM & True & AE & EM \\ 
      \hline
      M & 1 & 0.5 & 0.27 & 0.51 & 1.69 & 0.62 & 1.90 & 0.92 & 0.67 & 0.89 \\ 
		M & 1 & 1 & 0.22 & 0.36 & 0.37 & 0.59 & 0.70 & 0.94 & 0.67 & 0.72 \\ 
		M & 1 & 1.5 & 0.19 & 0.33 & 0.26 & 0.60 & 0.61 & 0.94 & 0.69 & 0.72 \\ 
		M & 1 & 2 & 0.16 & 0.33 & 0.29 & 0.63 & 0.66 & 0.94 & 0.68 & 0.73 \\ 
		M & 10 & 0.5 & 0.81 & 1.73 & 2.29 & 0.59 & 0.82 & 0.90 & 0.59 & 0.62 \\ 
		M & 10 & 1 & 0.91 & 1.70 & 1.33 & 0.51 & 0.57 & 0.95 & 0.60 & 0.67 \\ 
		M & 10 & 1.5 & 0.83 & 1.70 & 1.26 & 0.54 & 0.59 & 0.95 & 0.60 & 0.68 \\ 
		M & 10 & 2 & 0.75 & 1.68 & 1.29 & 0.57 & 0.63 & 0.95 & 0.61 & 0.68 \\ 
		M & 100 & 0.5 & 1.86 & 9.45 & 6.95 & 0.68 & 0.68 & 0.84 & 0.46 & 0.48 \\ 
		M & 100 & 1 & 2.63 & 9.38 & 5.90 & 0.51 & 0.51 & 0.95 & 0.45 & 0.53 \\ 
		M & 100 & 1.5 & 2.83 & 9.41 & 5.89 & 0.48 & 0.50 & 0.95 & 0.46 & 0.54 \\ 
		M & 100 & 2 & 2.76 & 9.50 & 6.01 & 0.48 & 0.51 & 0.95 & 0.46 & 0.54 \\ 
		\hline
		P & 1 & 0.5 & 0.65 & 1.05 & 2.41 & 0.69 & 1.76 & 0.92 & 0.68 & 0.89 \\ 
		P & 1 & 1 & 0.22 & 0.34 & 0.86 & 0.65 & 0.97 & 0.94 & 0.71 & 0.78 \\ 
		P & 1 & 1.5 & 0.22 & 0.33 & 0.27 & 0.62 & 0.62 & 0.95 & 0.72 & 0.75 \\ 
		P & 1 & 2 & 0.17 & 0.33 & 0.32 & 0.66 & 0.73 & 0.95 & 0.72 & 0.75 \\ 
		P & 10 & 0.5 & 1.21 & 2.30 & 3.18 & 0.60 & 0.85 & 0.90 & 0.59 & 0.62 \\ 
		P & 10 & 1 & 1.14 & 1.92 & 1.71 & 0.53 & 0.60 & 0.95 & 0.60 & 0.65 \\ 
		P & 10 & 1.5 & 1.04 & 1.81 & 1.35 & 0.57 & 0.59 & 0.95 & 0.62 & 0.68 \\ 
		P & 10 & 2 & 0.81 & 1.87 & 1.36 & 0.60 & 0.63 & 0.95 & 0.62 & 0.68 \\ 
		P & 100 & 0.5 & 2.78 & 12.81 & 8.90 & 0.65 & 0.75 & 0.84 & 0.44 & 0.50 \\ 
		P & 100 & 1 & 3.63 & 12.34 & 7.64 & 0.48 & 0.54 & 0.90 & 0.45 & 0.56 \\ 
		P & 100 & 1.5 & 3.38 & 11.37 & 6.99 & 0.48 & 0.53 & 0.92 & 0.46 & 0.57 \\ 
		P & 100 & 2 & 2.81 & 11.67 & 6.90 & 0.48 & 0.54 & 0.93 & 0.47 & 0.58 \\
		\hline 
		N & 1 & 0.5 & 0.54 & 1.00 & 2.32 & 0.69 & 1.77 & 0.93 & 0.67 & 0.89 \\ 
		N & 1 & 1 & 0.22 & 0.34 & 0.88 & 0.65 & 0.97 & 0.94 & 0.71 & 0.78 \\ 
		N & 1 & 1.5 & 0.22 & 0.32 & 0.26 & 0.63 & 0.62 & 0.95 & 0.72 & 0.75 \\ 
		N & 1 & 2 & 0.17 & 0.33 & 0.32 & 0.67 & 0.74 & 0.95 & 0.72 & 0.75 \\ 
		N & 10 & 0.5 & 1.21 & 2.33 & 3.13 & 0.60 & 0.85 & 0.90 & 0.58 & 0.61 \\ 
		N & 10 & 1 & 1.13 & 1.93 & 1.74 & 0.56 & 0.61 & 0.94 & 0.61 & 0.64 \\ 
		N & 10 & 1.5 & 1.02 & 1.84 & 1.36 & 0.58 & 0.60 & 0.95 & 0.62 & 0.68 \\ 
		N & 10 & 2 & 0.77 & 1.85 & 1.36 & 0.62 & 0.65 & 0.95 & 0.63 & 0.68 \\ 
		N & 100 & 0.5 & 2.84 & 12.96 & 8.89 & 0.65 & 0.74 & 0.85 & 0.45 & 0.51 \\ 
		N & 100 & 1 & 3.65 & 12.36 & 7.63 & 0.49 & 0.54 & 0.90 & 0.46 & 0.56 \\ 
		N & 100 & 1.5 & 3.37 & 12.04 & 6.96 & 0.48 & 0.53 & 0.92 & 0.46 & 0.57 \\ 
		N & 100 & 2 & 2.93 & 11.85 & 6.96 & 0.49 & 0.54 & 0.93 & 0.47 & 0.58 \\ 
		\hline
  \end{tabular}
\end{table}

\begin{figure}[htb]
\begin{center}
	\includegraphics[width=0.6\columnwidth, angle=270]{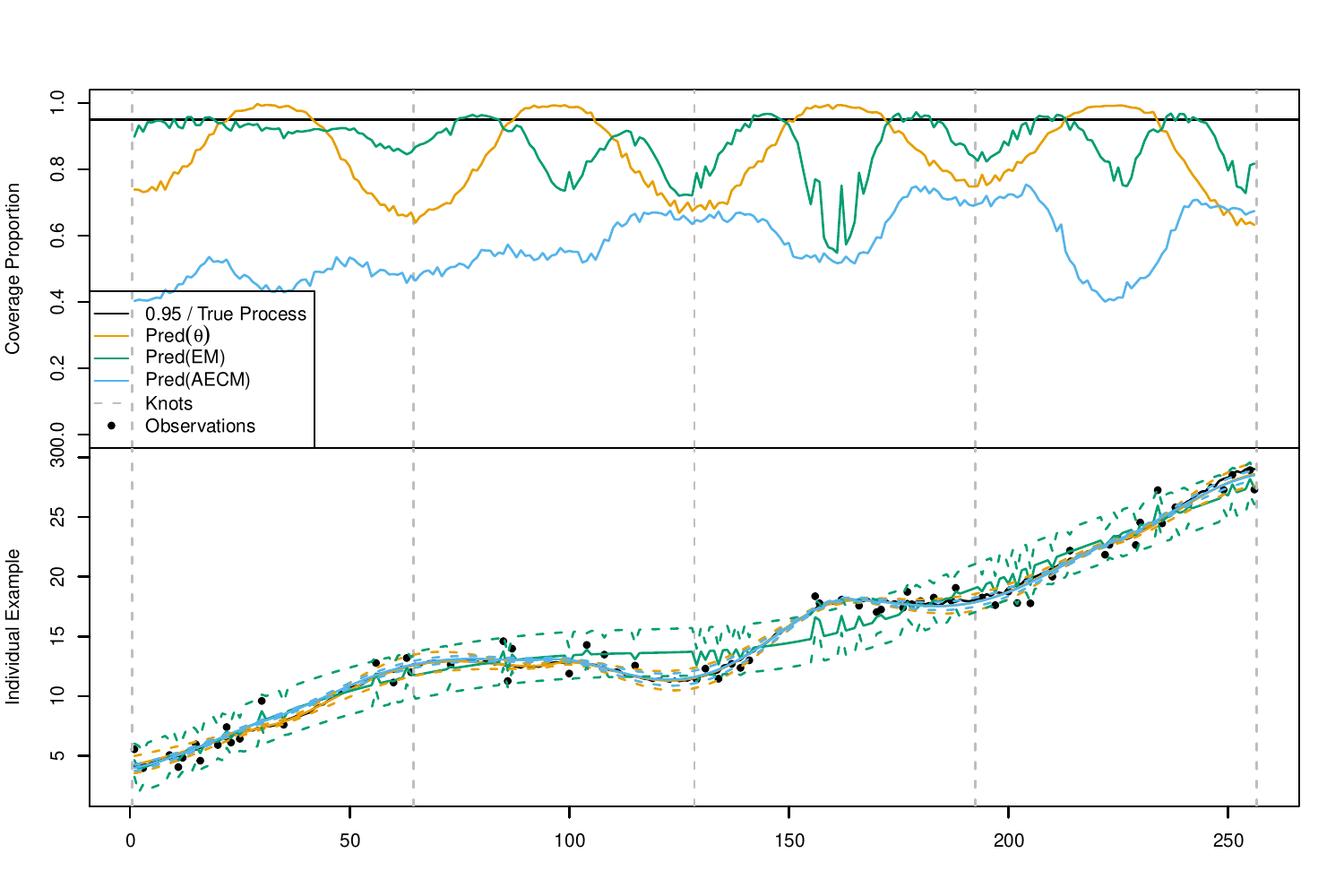}
	\caption{Simulated 1-dimensional spatial field where $b=0.5$, $\sigma_{\delta}^2 = 0.01$, 
		$\sigma_{\epsilon}^2 = 1$, $\bK \sim W_p(\ldots)$, and observations are randomly dispersed. 
		The top plot shows the PIC, computed over all 1000
                simulations, while the bottom plot depicts 
		a single realization.}
	\label{Simplot}
\end{center}
\end{figure}

Clearly, knowing the true parameter values improves prediction, as its
MSPE is always  lowest. If the measurement error is small enough and
the true $b \le 1$, the AECM algorithm  
improves MSPE compared to the EM approach. However, for
$\sigma_{\epsilon}^2 \ge 10$, holding $b=1.5$ fixed 
provides the most accurate prediction. The KSE is often
underestimated, with the AECM method  
exacerbating the issue. However, when the true $b=0.5$ and
$\sigma_{\epsilon}^2 = 1$, the EM  
approach severely overestimates KSE. The lower KSE also has the
consequence of lowering the PIC,  
as narrower confidence bands will inevitably miss the true values at a
higher rate.  
However, the anticipated improved performance using AECM when $b=0.5$ and $\sigma_{\epsilon}^2 = 1$ 
is absent. To elucidate this pattern, a single realization from 
our simulation with prediction regions is provided in Figure \ref{Simplot}, with the prediction 
interval coverage (PIC) at each location included to facilitate comparison.
The PIC is exclusively closer to 0.95 for the EM method across locations (top plot), but 
as a plot of the actual process (bottom) accentuates, the EM method misses clear signals in the 
data by over-smoothing predictions. Overestimating KSE, in this case, allows the prediction intervals 
to include the true values, however, it is an inaccurate representation of the process.

In summary, the results of our experiments show that estimating the bandwidth constant in the 
local bisquare basis functions by means of the AECM algorithm can be advantageous to both 
individual parameter estimation and prediction when the true $b$ is small, particularly when the 
measurement error is also low. This combination produces obvious large-scale trends in the 
data that is not captured by the mean structure. Clearly, fitting a more complex model for the 
mean, such as a higher order polynomial, would also allow the flexibility necessary to fit such 
models. The decision of whether or not to estimate $b$ is then analogous to a common statistical 
concern in mixed models of whether to treat an effect as fixed (mean structure) or random 
(spatial process), which depends on the context and desired interpretation.
As expected, our results as $b$ approaches 1.5 confirm that the extra
parameter estimation is unnecessary,  
however the potential negative effects on prediction accuracy are
fairly minimal. In terms  
of overall model fit, as measured by KL divergence, estimating $b$ by
means of the AECM algorithm  
still shows improvement over the standard approach of fixing
$b=1.5$. Having gained these insights from our simulation study, we
now apply our methodology to the National Climate Data Center (NCDC)
data of monthly  temperatures recorded across the continental United
States of America.

\section{Application: Predicting temperatures over the contiguous United States}  
\label{application}

We applied our methodology to temperature data recorded by 
the Cooperative Observer Program (COOP). Established in 1890, The COOP
is the  largest and oldest weather and climate observation network
\cite{coop2000}, recording weather information every 24 hours
from over 11,700 volunteer citizens and institutions. The
data are available online at the Uniform
Resource Locator 
\href{http://www.image.ucar.edu/Data/US.monthly.met/}{http://www.image.ucar.edu/Data/US.monthly.met/}. Note
that there is considerable variability in the 
number of weather  stations across the contiguous United States (US)
for which measurements are available at any point of time as well as
in the number of variables recorded (monthly maximum temperature, 
minimum temperature, precipitation, etc.) at these time points.

Mean temperature recorded over a regular grid is an important summary in 
climate science \cite{johnsnychkakitteldaly2003}, as it is used to 
assess the value of climate-derived models by means of comparison. 
Inevitably, temperatures are not recorded for all locations on the 
predetermined regular grid. Thus, kriging is necessary to estimate
temperatures at the remaining unobserved locations. Also, since spatial 
dependence typically exhibits a high degree of smoothness with respect to 
temperature, this response should help determine if $b=1.5$ is a good
default smoothing parameter value. 

We restrict our attention to mean daily temperature readings for April
1990 in this paper. The COOP had daily minimum and maximum temperatures
recorded at 5,030 locations across the contiguous US -- these were
summarized into a single univariate measurement at each of these
locations in terms of an average of the mean monthly minimum 
and maximum temperatures. In this setup, note that performing kriging
requires iteratively solving a $5030 \times 5030$ linear system, a
task which would be computationally prohibitive in terms of CPU time
and memory. Thus,  efficient forms of kriging such as fixed rank
kriging, with parameter estimation achieved through either an EM or
AECM approach are necessary.  
Note, also that this dataset is used simply for illustration, our
method can easily be applied to larger sample sizes, as it is scalable
with computational cost growing at a linear rate with the sample size.

\subsection{Known Covariates}
Temperature is known to be affected by geographic location and elevation. 
Since temperature cannot reasonably be modeled with a constant mean 
structure in the presence of these factors, a set of covariates is
necessary to reduce the remaining structure to a zero-mean spatial process
and zero-mean measurement error. Assuming linearity in the effects of
our covariates, we can model the non-zero mean structure with the
additive model $\bx(\bs,\bh_{\bs})'\bbeta = \beta_0 + \bg_{Elev}(\bh_{\bs})
\bbeta_{Elev} + \bg_{Lat}(\bs)\bbeta_{Lat} + \bg_{Lon}(\bs)\bbeta_{Lon}$ 
where $\bh_{\bs}$ is the elevation at location $\bs$.

\begin{figure}[htbp]
  \mbox{
    \subfloat[Cubic spline]{\includegraphics[width=.21\columnwidth,angle=270]{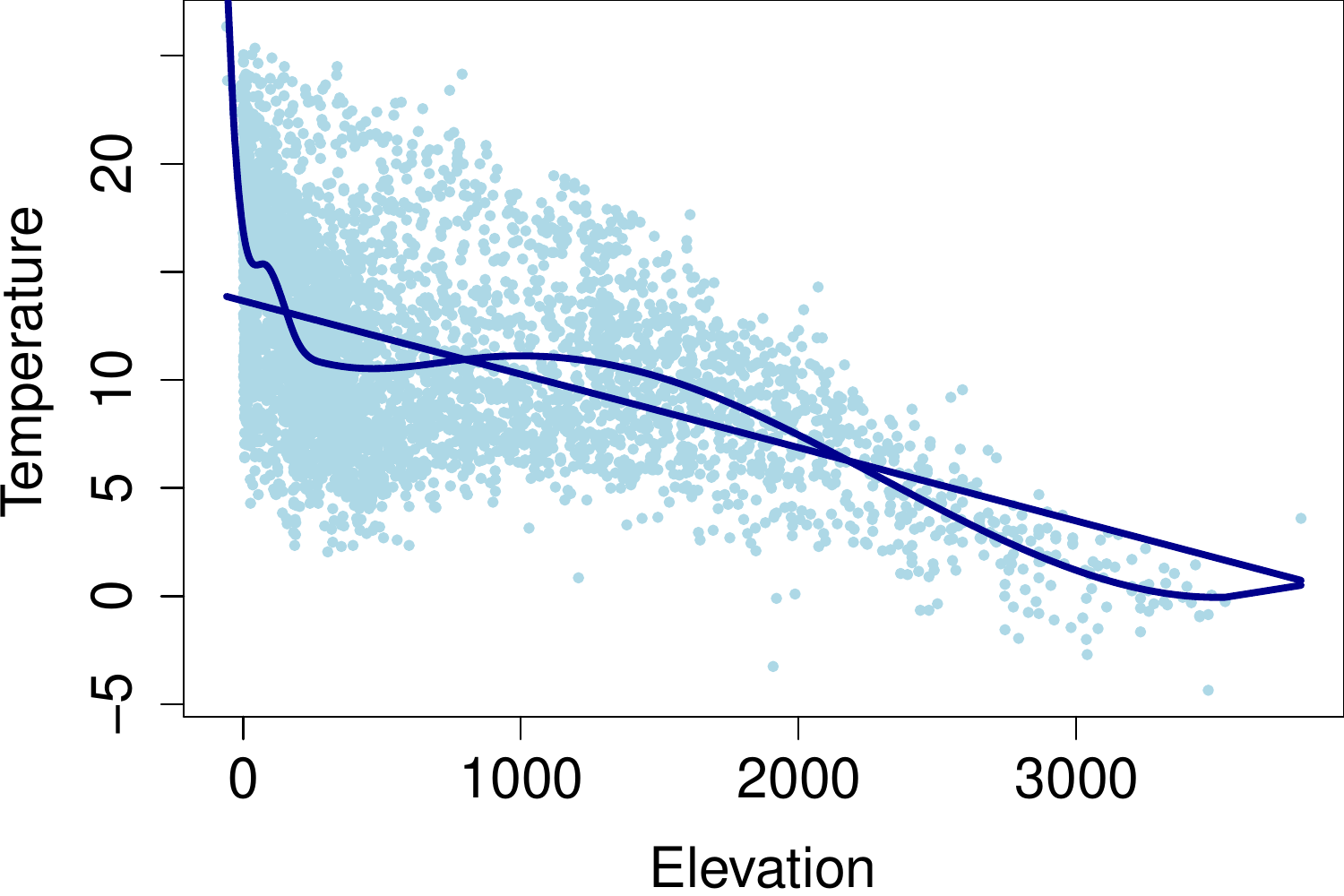}}
    \subfloat[Quadratic spline]{\includegraphics[width=.21\columnwidth,angle=270]{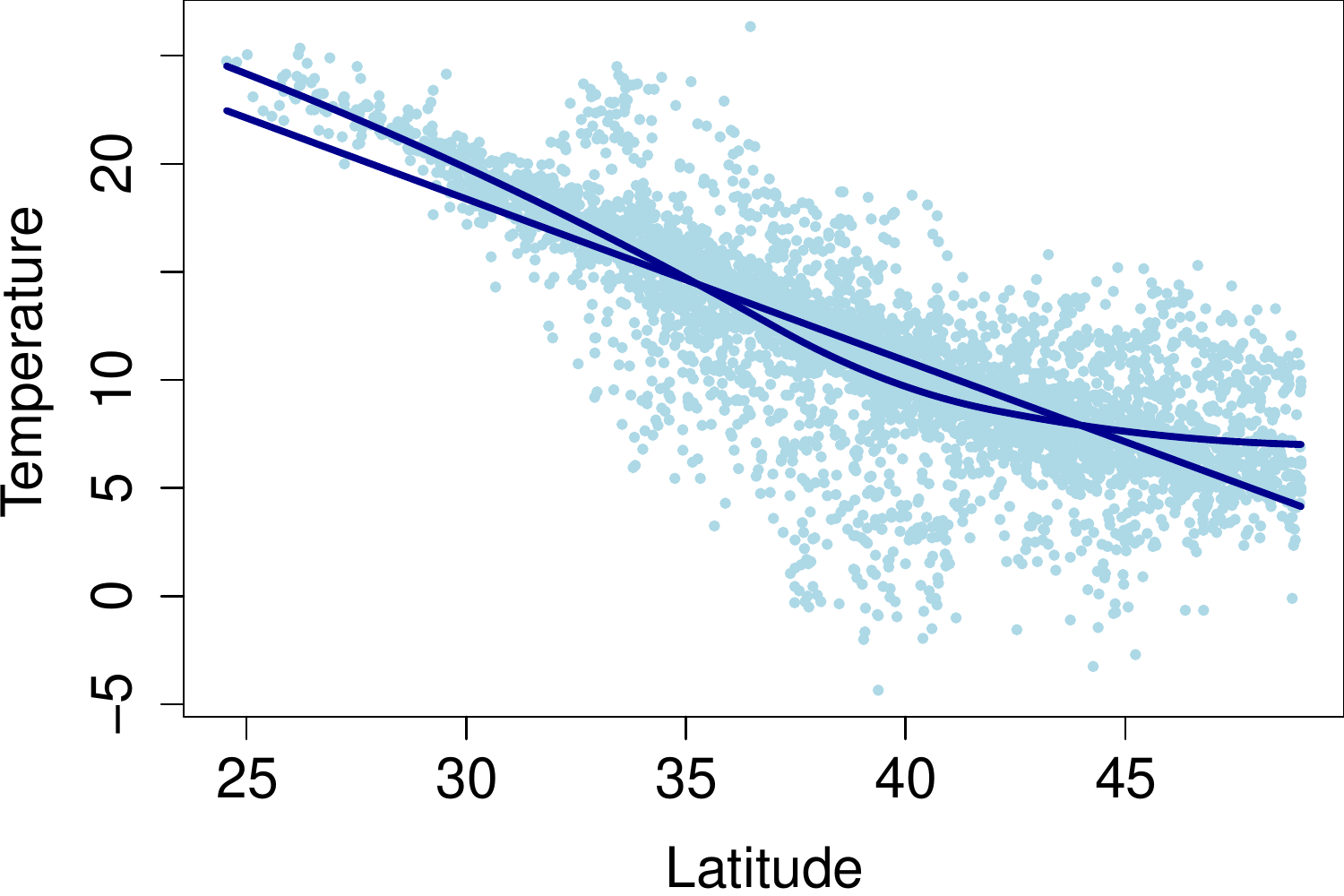}}
    \subfloat[Cubic spline]{\includegraphics[width=.21\columnwidth,angle=270]{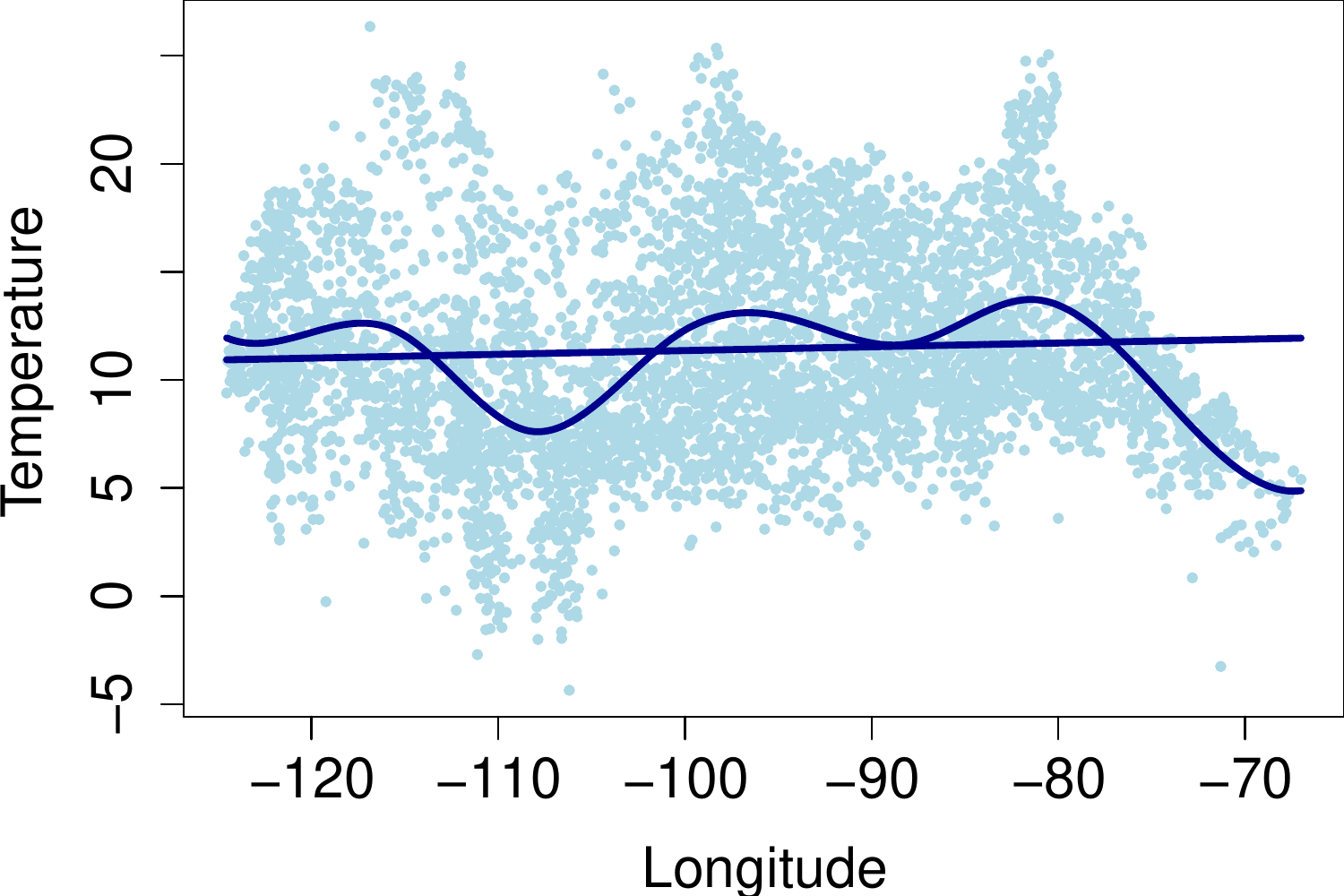}}}
\caption{(a) - (c) display monthly mean temperature plotted against one 
  of three covariates with linear fits and regression splines.}
\label{splines}
\end{figure}

\begin{figure}[htbp]
  \mbox{
    \subfloat[Full model]{\includegraphics[width=0.31\columnwidth,angle=270]{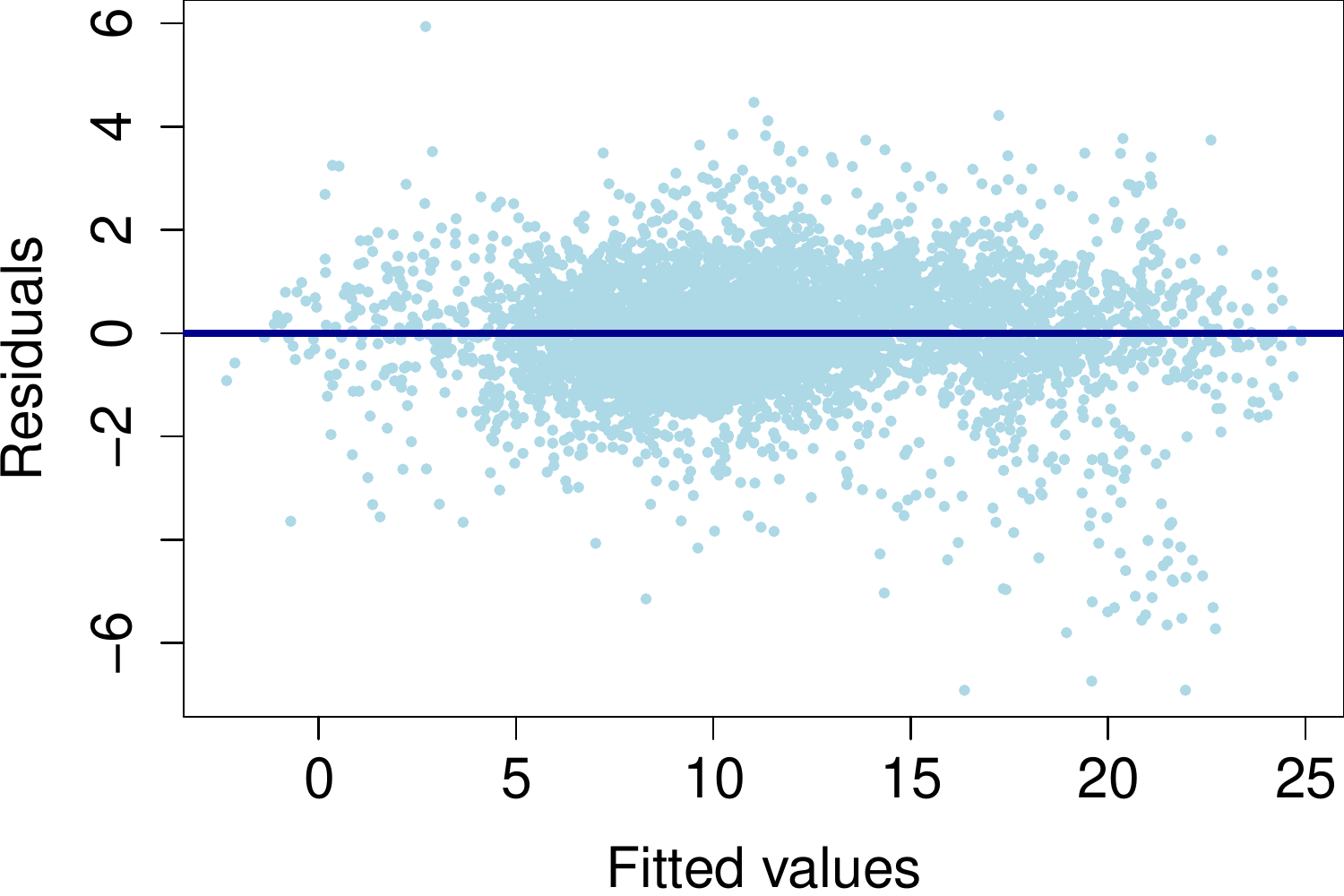}}
    \subfloat[Reduced model]{\includegraphics[width=0.31\columnwidth,angle=270]{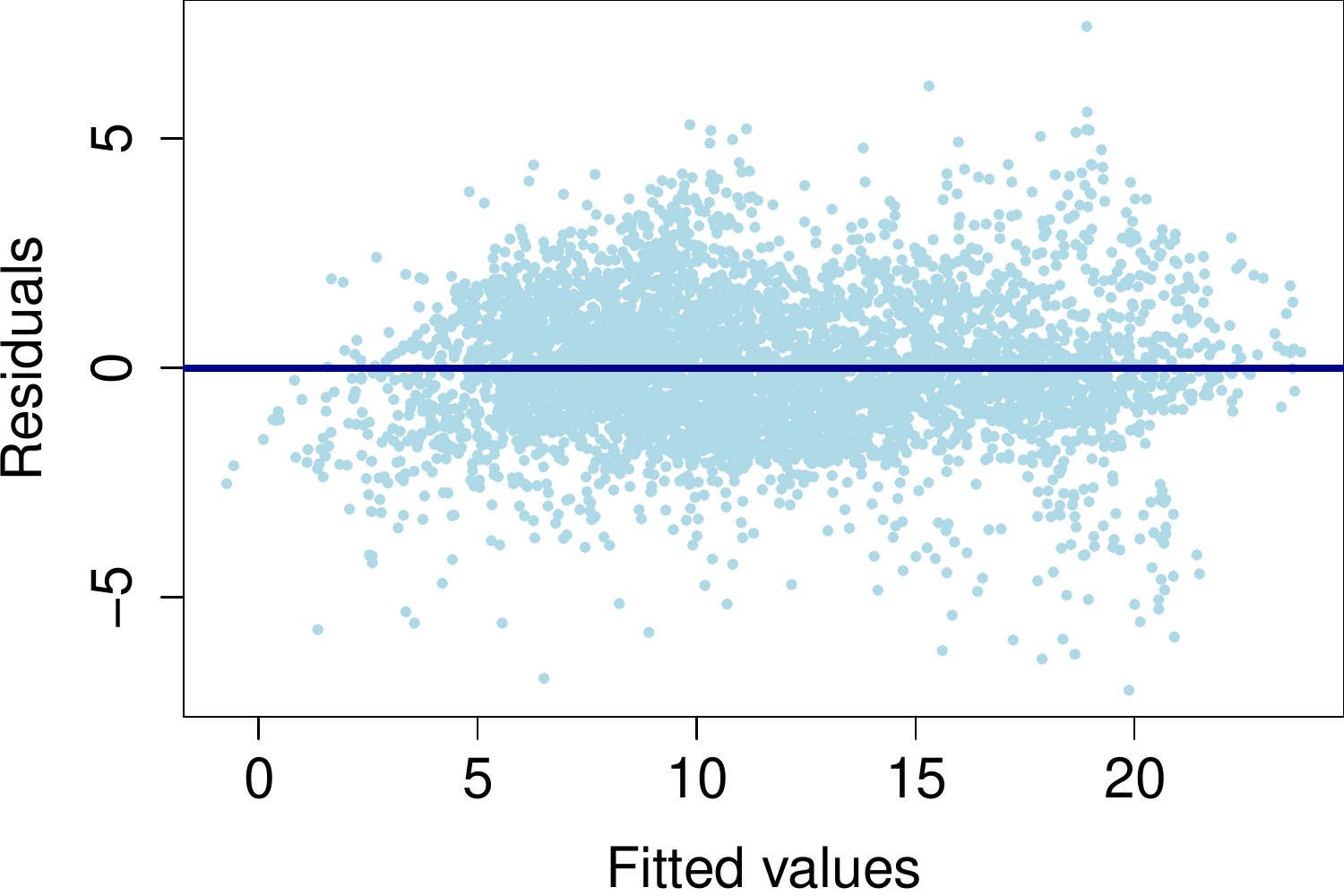}}}
  \caption{Residual plots based on a fit using (a) splines and (b) a linear 
    model.}
  \label{resids}
\end{figure}

The exact functional relationship between temperature and our covariates 
is unknown, however we expect temperature to decrease with increasing 
elevation and latitude. One option is to assume a simple linear mean 
structure for all covariates. However, this is likely an oversimplification 
and, hence, smoothing regression splines fit based on AIC were also used 
to model the mean structure. Specifically, $\bg_{Elev}(\bh_{\bs})$ 
required a cubic regression spline with nine degrees of freedom, 
$\bg_{Lat}(\bs)$ was modeled with a quadratic regression spline with four 
degrees of freedom, and  $\bg_{Lon}(\bs)$ fit best with a cubic 
regression spline with nine degrees of freedom. Since the AECM method showed 
the ability to accommodate irregular patterns in the data not captured by the 
mean structure, we are interested in a comparison between these ``reduced'' 
and ``full'' models. The model fits and their resulting residuals, are shown in 
Figures \ref{splines} and \ref{resids}, respectively. Although the relationship 
between temperature and each covariate appears decidedly non-linear, 
validation of model assumptions, by way of a residual plot, does not feature 
this violation. Visual assessment suggests that both mean structures 
are reasonable.

\subsection{Results and Analysis}
REML estimation was carried out by maximizing the restricted
log-likelihood \eqref{altlike} by the algorithm outlined in Section 
\ref{aecm} and by the iterative updating scheme of \eqref{M} obtained 
by the EM algorithm. 
Following the strategy proposed by \cite{kangcressieshi2010}, 
$\sigma_{\epsilon}^2 = 2.059$ was independently determined.
We selected $m=369$ knots representing 7.3\% of the original data set 
on a regular triangular grid within a space-filling framework and
tested our methodology on up to four levels of resolution. Distance 
matrices were obtained by using the function \texttt{rdist.earth}, which 
computes the Great circle distance between any two locations on the Earth.

The size of this dataset, despite making use of the fixed rank kriging 
equations, also warranted the use of more efficient software.
Functions were translated to \texttt{MATLAB} \cite{matlab2014}, which is 
optimized for matrix and vector algebra and can be easily
parallelized. We note that a major burden 
of the AECM method is that every iteration during parameter
estimation costs three to four  
times that of the EM method and that subsequent iterations require
previous  results, so care had to be taken in parallelization.
However, despite the additional burden, recall that the main benefit of 
our AECM algorithm is additional parameter estimation.

We compared model performance (a) in terms of prediction by MSPE through 
five-fold cross-validation and (b) in terms of efficiency, summarized
by the elapsed  
time for estimation, in minutes (see Table 2). 
All reported computations was performed using a standard desktop computer with 16GB of RAM 
and a 64-bit Intel quad-core i7 processor. Also included in the table are 
parameter estimates for $b$, the range of $\bK$, and 
$\sigma_{\delta}^2$ and the corresponding evaluated log-likelihood.

\begin{table}
	\caption{\label{app_results} Prediction and estimation results
          from both the EM and AECM algorithms, using 1 to 4
          resolutions ($\varrho$) and the full (top block) or reduced (bottom
          block) model on the US 
          temperature dataset. The model column provides both the model and resolution. R($\hat{\bK}$) represents the range of values in the $\hat{\bK}$ matrix.}
        
	\begin{tabular}{ll|rr|rrrr}
	  \hline
			$\varrho$ & Method & MSPE & Min. & $\hat{b}$ & R($\hat{\bK}$) & 1E3 $\widehat{\sigma_{\delta}^2}$ & $\ell(\hat{\btheta})$ \\
			\hline
			 1 & EM & 0.771 & 14.6 & NA & 158.64 & 0.0046 & -7215.7 \\
			 1 & AECM & 0.772 & 26.2 & 1.51 & 155.26 & 0.0046 & -7215.8 \\
			 2 & EM & 0.767 & 16.8 & NA & 210.11 & 0.0047 & -7213.0 \\
			 2 & AECM & 0.767 & 27.3 & 1.47 & 203.06 & 0.0047 & -7212.6 \\
			 3 & EM & 0.768 & 17.1 & NA & 318.34 & 0.0046 & -7214.7 \\
			 3 & AECM & 0.769 & 28.3 & 1.47 & 318.27 & 0.0046 & -7215.2 \\
			 4 & EM & 0.768 & 19.0 & NA & 326.18 & 0.0046 & -7214.9 \\
			 4 & AECM & 0.770 & 32.5 & 1.48 & 336.32 & 0.0046 & -7216.3 \\
			\hline\hline
			 1 & EM & 0.805 & 23.0 & NA & 537.85 & 0.0030 & -7257.4 \\
			 1 & AECM & 0.790 & 37.0 & 1.89 & 297.90 & 0.0034 & -7244.9 \\
			 2 & EM & 0.795 & 25.8 & NA & 507.49 & 0.0031 & -7245.3 \\
			 2 & AECM & 0.792 & 39.7 & 1.55 & 603.01 & 0.0033 & -7242.1 \\
			 3 & EM & 0.793 & 30.3 & NA & 575.08 & 0.0030 & -7244.8 \\
			 3 & AECM & 0.795 & 47.8 & 1.27 & 372.84 & 0.0033 & -7245.4 \\
			 4 & EM & 0.794 & 27.4 & NA & 617.49 & 0.0030 & -7245.6 \\
			 4 & AECM & 0.798 & 41.1 & 1.28 & 354.57 & 0.0032 & -7245.1 \\
			\hline
		\end{tabular}
\end{table}

The performance measures show that, as expected, predictions stemming
from the full model provide lower MSPEs.  
When using the full model, 2 resolutions minimized the MSPE, with the AECM 
method barely outperforming the EM method. Three and two resolutions were 
optimal when using the reduced linear model for the EM and AECM methods, 
respectively. Here again, AECM produced lower prediction errors. 
The largest discrepancy occurred for the simplest model, reduced with
only one resolution, where the AECM approach substantially decreased
MSPE, relative to the EM.
Although the AECM algorithm is not optimal across the board for every
number of resolutions,  these results also indicate that prediction
error is not substantially increased.

The EM method is, by design, a more efficient algorithm and the timing 
results substantiate that fact. Although the AECM method does require 
additional computation time, the increase observed was only between 
50\% and 79\%. Considering the amount of communication required between 
workers (necessary after each iteration), \texttt{MATLAB} manages this 
extra burden surprisingly well. Also note that the relative increase was 
lower across all resolutions when assuming the reduced model.

For the full model, parameter estimates for $\sigma_{\delta}^2$ were essentially 
identical between the two methods and estimation of $b$ validated the 
reasonable choice of bandwidth constant being 1.5, as mentioned by 
\cite{cressiejohannesson2008}. However, the resolution that both 
maximized the evaluated log-likelihood and minimized the MSPE also 
provided an estimate of $b$ with the largest deviation from 1.5. The 
reduced model produced estimates of $b$ as high as 1.890, with the optimal 
resolution, as measured by evaluated log-likelihood, estimating $b$ at 1.546. 
All values of $\hat{\sigma}_{\delta}^2$ for the reduced model are lower relative to 
the full model, but the decrease is consistently less for the AECM method. If we 
assume that the full model is a more appropriate description of our data, then 
this provides evidence that the AECM method counteracts the effects of model 
misspecification by allowing the range of the basis functions, or bandwidth, to 
vary. The range of $\hat{\bK}$ using the AECM method also shows greater 
agreement between analogous models for almost every resolution, further 
corroborating this hypothesis. Complete image plots of $\hat{\bK}$ are 
provided in Section S-1.2 of the Supplementary Materials.

\begin{figure}[htbp]
	\begin{center}
          \mbox{
	    \subfloat[EM]{\includegraphics[width=0.25\columnwidth,angle=270]{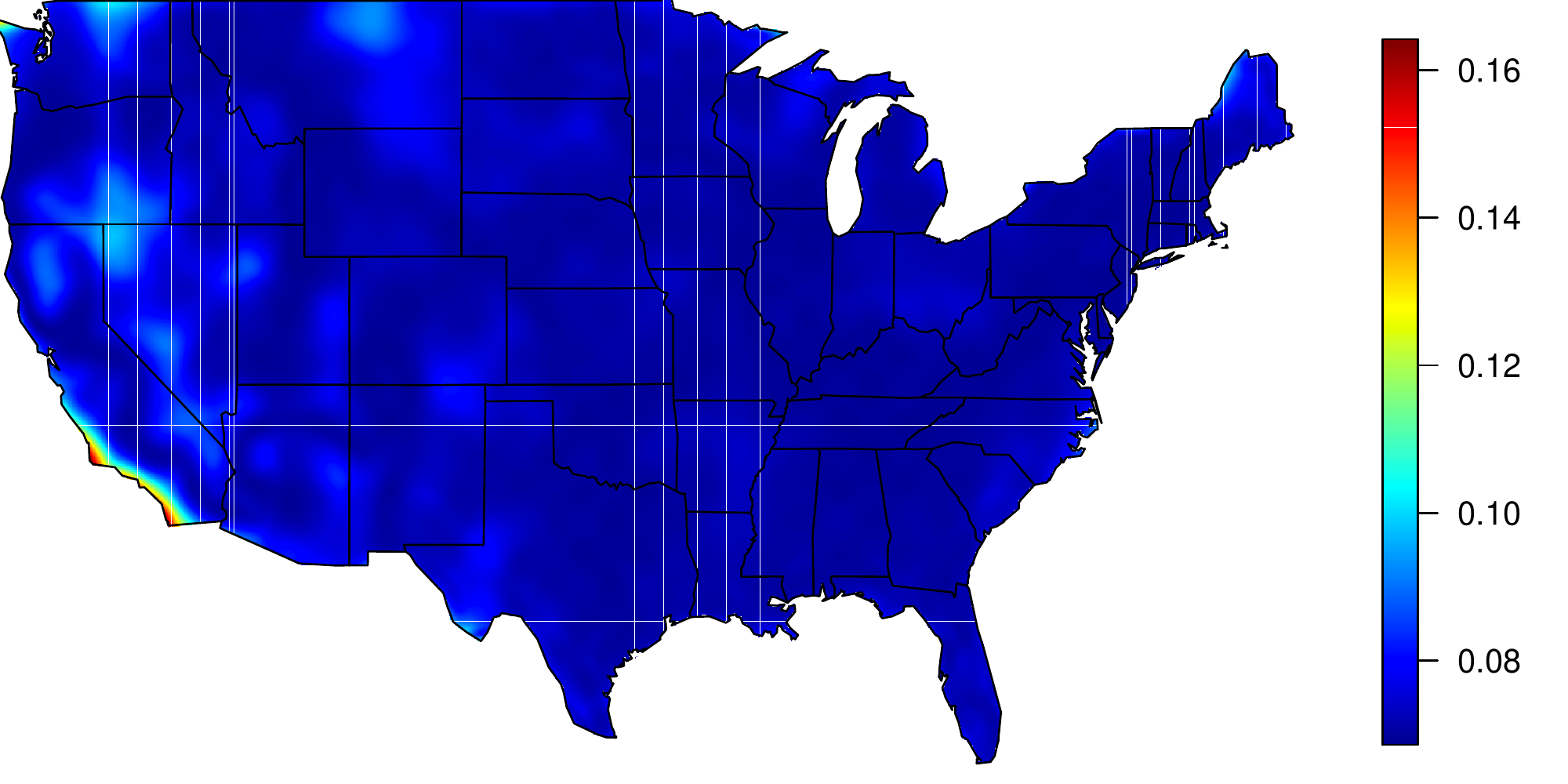}\label{USkseEM}}
	  \subfloat[AECM]{\includegraphics[width=0.25\columnwidth,angle=270]{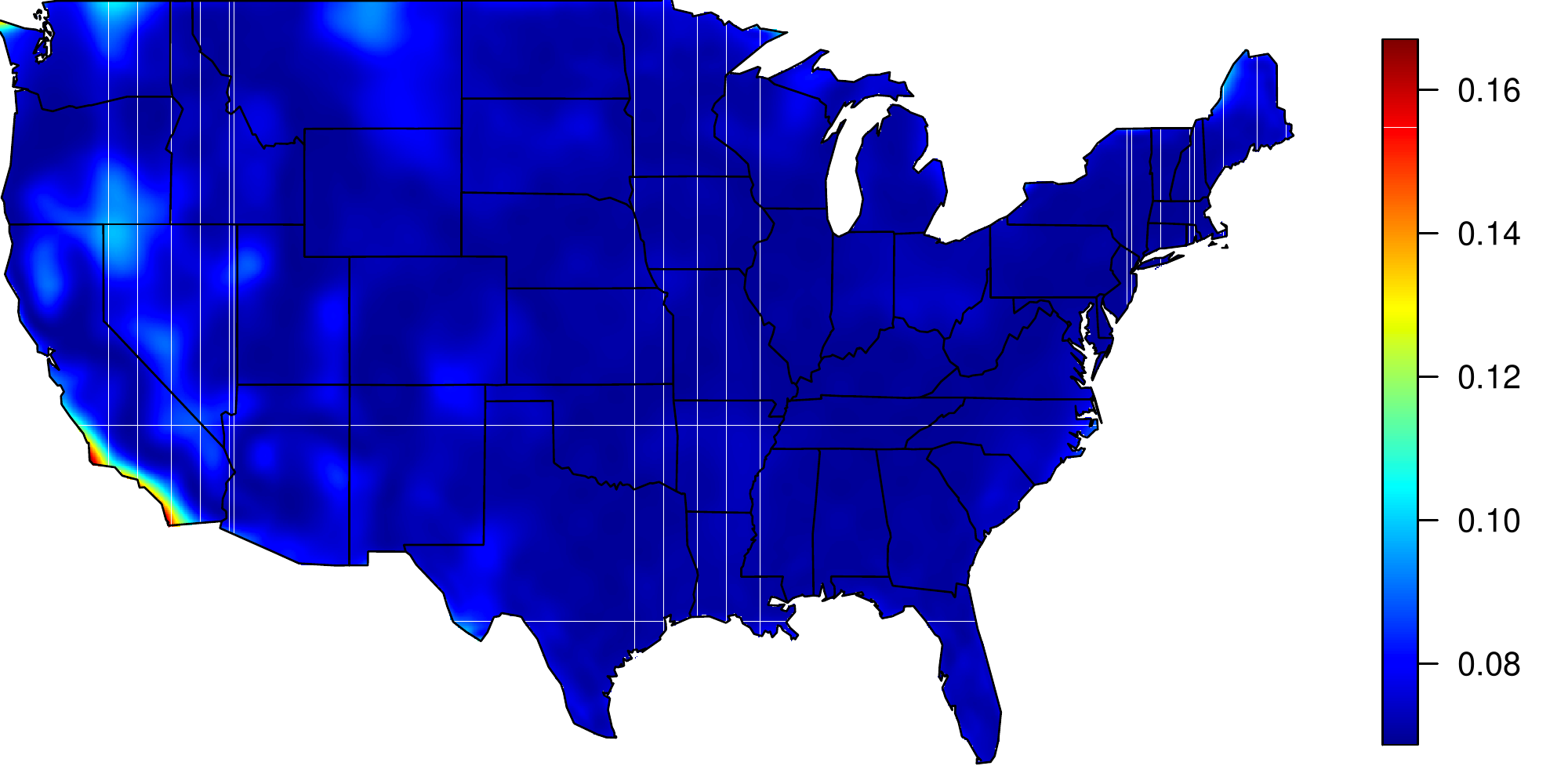}\label{USkseAECM}}}
	\end{center}
	\caption{Kriging standard errors assuming the full model with 2 resolutions 
		for both the (a) EM and (b) AECM methods.}
	\label{USkse}
\end{figure}
For the remainder of this section, we assume that the full model with 
two resolutions is our best model and, consequently, we focus on those 
results. Plots of the KSEs using both EM and AECM methods are provided 
in Figure \ref{USkse}. Both algorithms produce extremely similar results, 
however the scales are slightly shifted. The range of KSE values for the 
EM method is between 0.0688 and 0.1640, whereas the AECM method produces 
values between 0.0689 and 0.1669. Considering the evidence we observed 
that the fixed rank kriging approach underestimates KSE when $b \approx 
1.5$, we are inclined to believe that the estimates of KSE resulting from 
the AECM method are likely closer to the truth. Other than this difference, 
the pattern of KSE is fairly consistent between methods across the entire 
contiguous US, with pockets of higher variability existing in 
the western states, particularly along the southern coast line of California.

\begin{figure}[htb]
  \begin{center}
    \mbox{
      \subfloat[Observations]{\includegraphics[width=0.5\columnwidth]{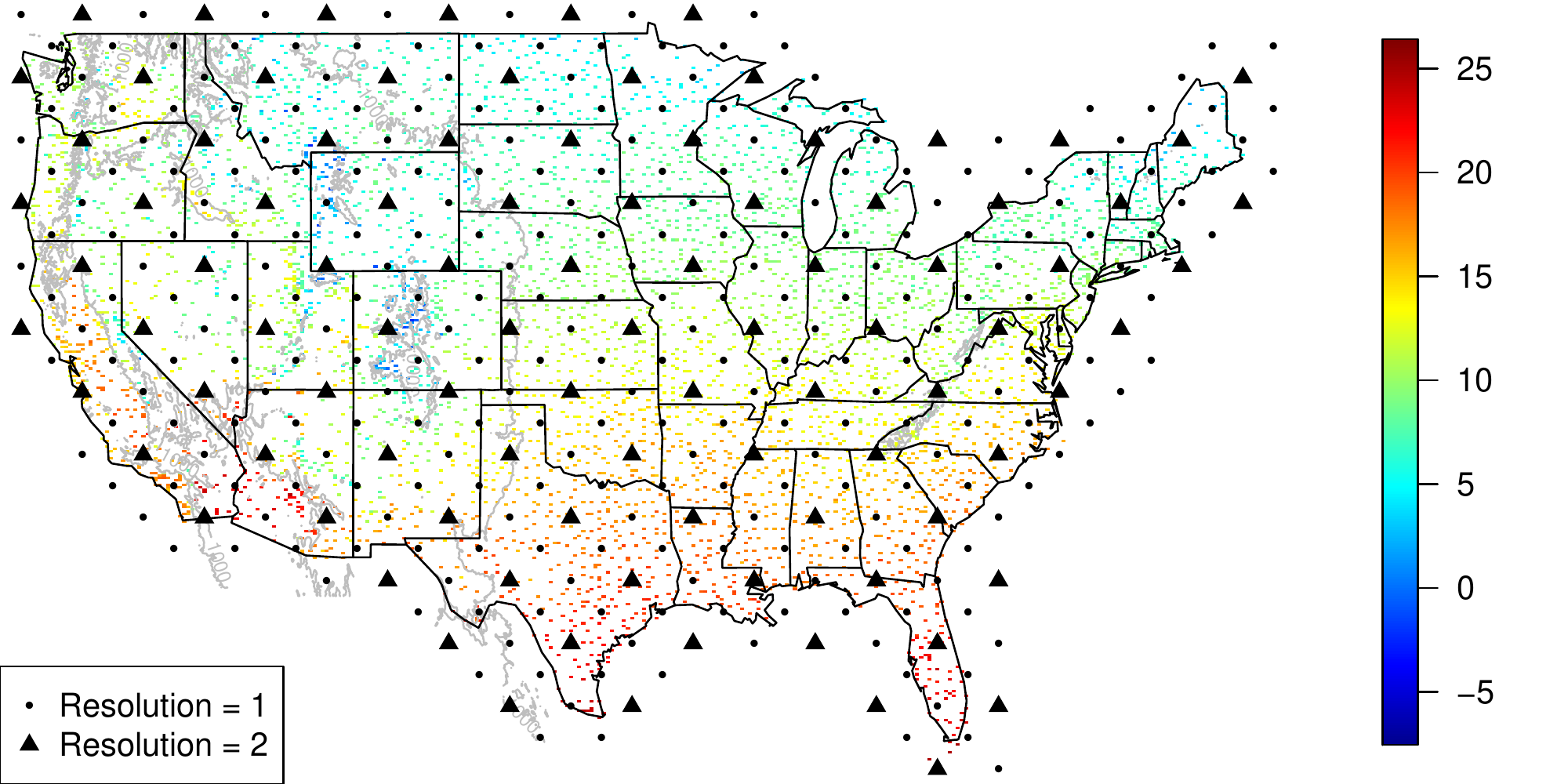}\label{USdata}}
      \subfloat[Mean Structure]
                {\includegraphics[width=0.5\columnwidth]{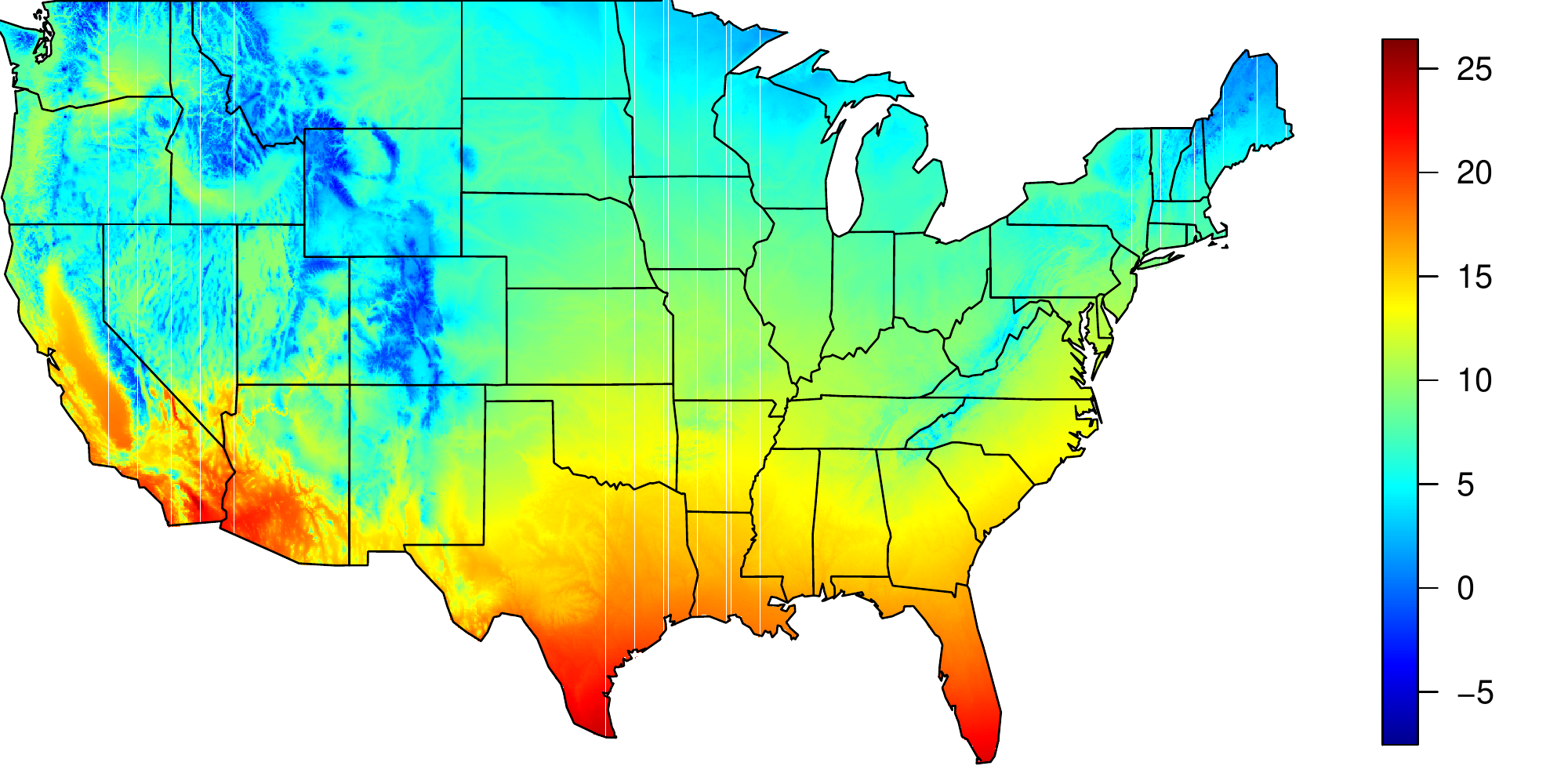}\label{USmu}}}
    \mbox{
      \subfloat[Spatial Dependence] {\includegraphics[width=0.5\columnwidth]{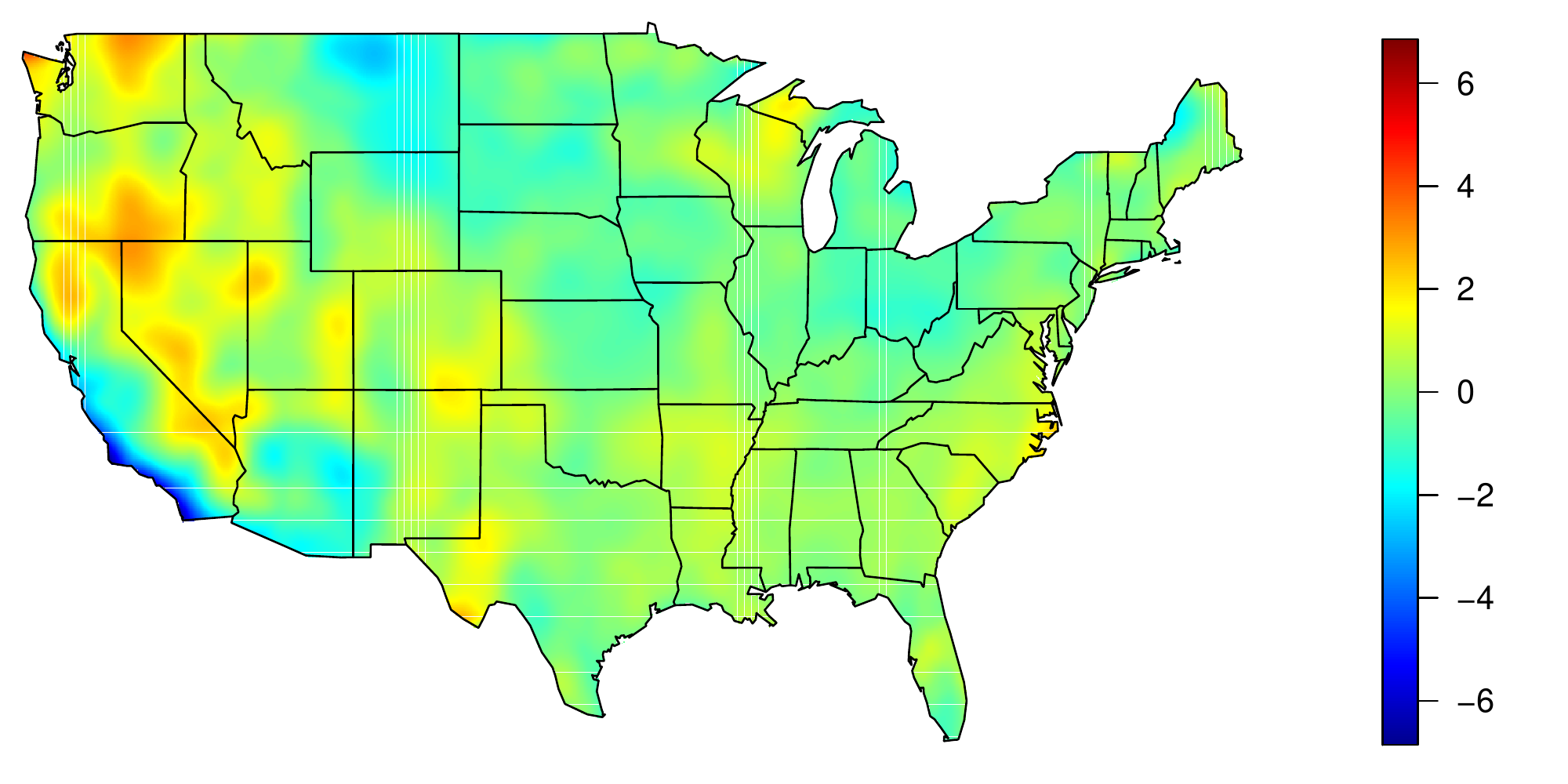}\label{USke}}
      \subfloat[Kriging Estimates] {\includegraphics[width=0.5\columnwidth]{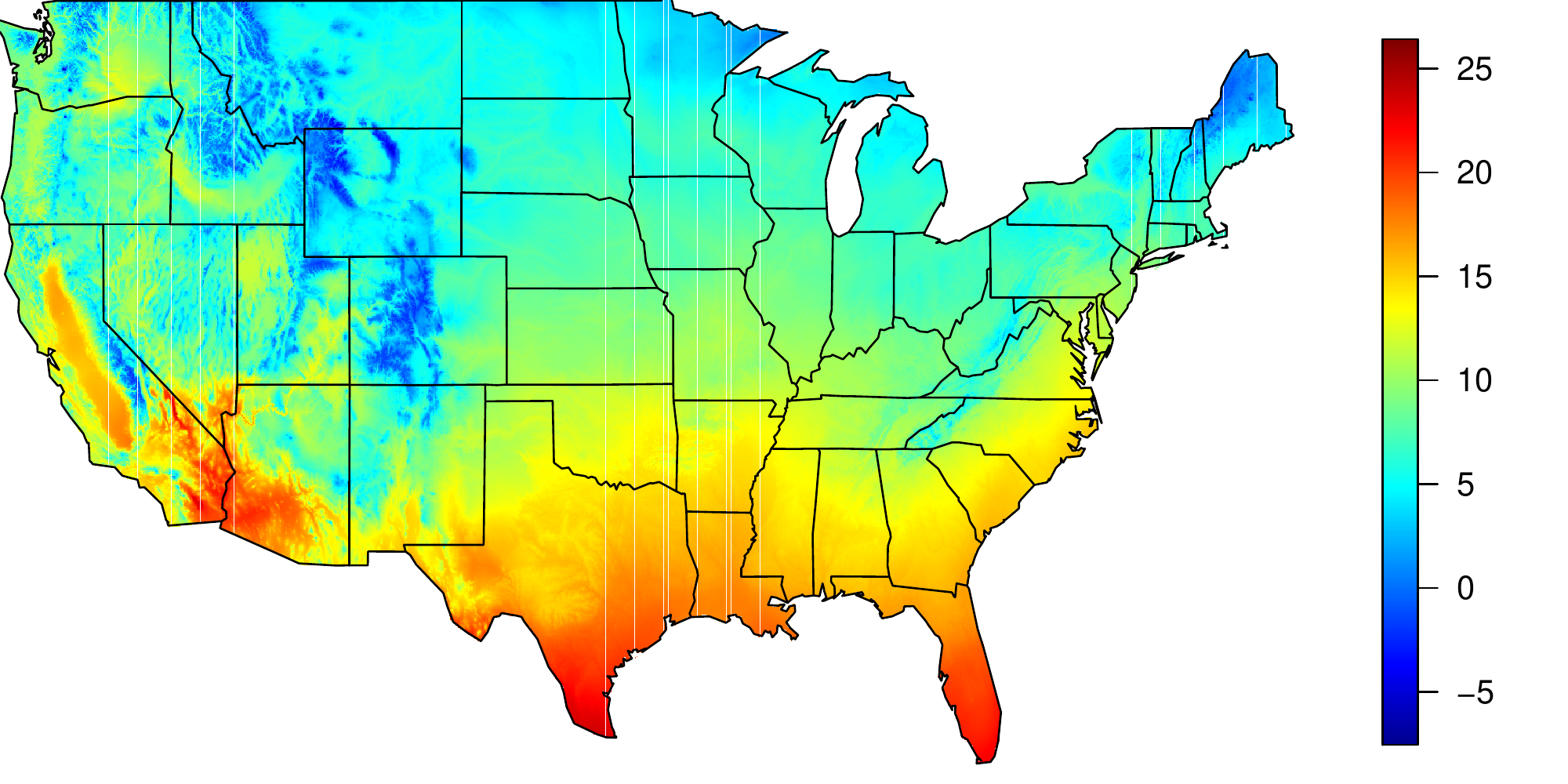}\label{USadd}}}
	\end{center}
	\caption{Mean temperatures in the US in April 1990; (a) displays the 5030 
		observations and 369 knots (black circles and triangles) at two resolutions, 
		(b) shows the portion of the predicted values explained by the mean structure, 
		(c) represents the portion of the predicted values explained by the spatial dependence, 
		and (d) depicts the complete additive kriging estimates.} 
	\label{USpred}
\end{figure}

Figure \ref{USpred} displays the data with superimposed knots
(Figure~\ref{USpred}a) for two resolutions 
and the corresponding spatially predicted field using the AECM method. The predictions 
are decomposed into three pieces: the mean structure ($\bX(\bs_0)\hat\bbeta$), 
the spatial dependence ($\bC(\bs_0) \bSigma^{-1}(\by - \bX\hat\bbeta)$), and the 
complete additive kriging estimates ($\hat\by(\bs_0) = \bX(\bs_0)\hat\bbeta + 
\bC(\bs_0) \bSigma^{-1}(\by - \bX\hat\bbeta)$).
It is clear from Figure \ref{USmu} that the covariates are critical to 
temperature prediction, since the mean structure reflects the general trends of 
the temperature in April in the contiguous US. In particular, the
image depicts the negative relationship that exists between 
elevation/latitude and temperature. However, Figure \ref{USke} shows that 
certain geographically related aspects of temperature are not easily described 
by a linear function of location. Mirroring patterns of high KSE, the spatially 
dependent components show warmer than expected areas in Washington, Oregon, 
California, and along most Nevada borders. Cooler than expected regions occur 
mainly along the southern coast line of California and in northern Montana. 
The cooler California coast line relative to location can be explained by the 
California Current which transports northern waters southward, cooling the 
ocean temperature near California and, consequently, its land temperature as 
well \cite{hickey_1979}. The resulting complete kriging 
estimates are found in Figure \ref{USadd}. The linear combination of the mean 
structure and the spatial dependence provide predictions with anticipated 
patterns, the mild climate relative to latitude that exists for the 
Pacific states being most evident. Thus, the results of applying our 
methodology on this dataset are explained by climate science and very
encouraging.


\section{Discussion}
\label{discussion}

This paper develops methodology to estimate a continuous tuning parameter in 
the SME model, analogous to the range parameter. The methodology maintains 
the ability to model severe nonstationarity in an efficient manner, without 
loss of prediction accuracy, thus  making it scalable to massive data
sets. This is accomplished using the SME model and fixed rank 
formulas developed by \cite{cressiejohannesson2008} and the EM
algorithm to obtain ML estimates of the parameters in the variance 
structure, suggested by \cite{katzfusscressie2011}. In addition  
to estimating variance components associated with the knots and
fine-scale variation ($\bK$ and $\sigma_{\delta}^2$), we estimate the
range parameter  in the basis functions by using the AECM algorithm. 
By making use of simple parallel processing options in \texttt{MATLAB}, 
we minimize the computational burden of additional parameter estimation.

Estimation of $b$ is valuable because knowledge of the range parameter in a 
spatial process can be of scientific interest. Our argument is that, 
in addition to knowledge gained about the spatial process, estimating $b$ can 
improve estimation of other parameters. By estimating $b$, we infuse flexibility 
into the model that allows the basis functions to describe the optimal range 
of spatial dependence. This flexibility has even been shown to reduce MSPE 
under certain situations when compared to its fixed $\bS$ counterpart. 

Although $\bK$ is completely flexible, outside of standard
requirements  
for covariance matrices, the locations of knots and the choice of basis 
function impose certain assumptions about the spatial dependence. 
The nonstationarity of $\bK$ is unconstrained, however, the pattern of 
nonstationarity in the full covariance $\bSigma$ is dependent on knot 
location. Also, in choosing the local bisquare basis function, we are assuming 
that the area of spatial dependence is circular, with diameter controlled by $b$. 
However, if an anisotropic pattern exists in the data, it would be straightforward 
to replace the local bisquare basis function with something more suitable. The 
current restriction is that the direction and relative scale of the anisotropic 
behavior would need to be known or estimated outside of the AECM algorithm.
Interesting directions for future work, thus, would include optimal knot location 
and an estimation procedure that can support multiple parameters in the basis 
functions. This would allow for more complex structures including, but 
not limited to, the estimation of the direction and scale of anisotropy 
and a varying range for each resolution ($b_l$). Thus, we see that
while we have addressed the issue of computationally practical
estimation and prediction in massive fields under the SME model,
several questions remain that are worthy of further attention.

\bibliographystyle{IEEEtran}
\bibliography{References}
\newpage
\renewcommand\thefigure{S-\arabic{figure}}
\renewcommand\thetable{S-\arabic{table}}
\renewcommand\thesection{S-\arabic{section}}
\renewcommand\thesubsection{S-\arabic{section}.\arabic{subsection}}

\section*{Supplementary Materials}
\subsection{Additional Tables and Figures}

\subsubsection{Experimental Evaluations}
A series of simulation experiments were conducted to explore the value of estimating 
a continuous component of the basis functions in fixed rank kriging. In particular, 
the bandwidth constant $b$ in the local bisquare function was estimated using the 
alternating expectation conditional maximization (AECM) algorithm and compared the 
the expectation maximization (EM) algorithm with a fixed $b=1.5$, as suggested by 
\citet{cressiejohannesson2008}.

Data were simulated from a one-dimensional spatial mixed effects (SME) model, 
similar to that used by \citep{katzfusscressie2011}. Observed locations, $n=64$, 
were selected from the complete spatial domain, $\mathcal{D} = \{1,\ldots,256\}$, 
either completely randomly or following a clustered random sample. 
Five knots were used, centered at 0.5, 64.5, 128.5, 192.5, and 256.5.

We used a simple linear mean structure ($\beta_0 = 5$ and $\beta_1=0.08$) and varied 
$\bK$ as either Mat\'ern, Wishart, or Wishart with strictly positive entries. 
Simulation values for the fine-scale variation parameter and the measurement error 
were, $\sigma_{\delta}^2 = \{0.01, 0.1, 1\})$ and $\sigma_{\epsilon}^2 = \{1, 10, 100\}$, 
respectively. Additional simulation inputs include the bandwidth constant 
($b = \{0.5, 1, 1.5, 2\}$), the number of resolutions ($l = 1$), and the weighting 
functions ($v_{\delta}(\cdot) = v_{\delta}(\cdot) = 1$). 

One thousand Gaussian random fields were simulated using the \texttt{R} 
\citep{cran2015} package {\tt fields} \citep{fields2015} for each combination of 
covariance type, parameter value, and sampling design. For each simulated field, we 
obtained MLEs using both the AECM algorithm (estimating $b$) and the EM algorithm 
(leaving $b=1.5$ fixed). From these estimates, we compared model fit 
quantified by Kullback-Leibler divergence and calculated the corresponding kriging 
estimates and standard errors at all locations within $\mD$.

\begin{table}
	\caption{\label{pred_M} Median predictions results, comparing mean square prediction error (MSPE), ratio of 
			median estimated versus true KSE, and prediction interval coverage (PIC) for a Mat\'ern covariance.}
     \scriptsize
		\begin{tabular}{llll|rrr|rr|rrr}
			\hline
			 & & & & & MSPE & & rKSE & & & PIC & \\
			$\sigma_{\delta}^2$ & $\sigma_{\epsilon}^2$ & $b$ & design & True & AECM & EM & AECM & EM & True & AECM & EM \\
			\hline
			0.01 & 1 & 0.5 & clus & 0.201 & 0.436 & 1.660 & 0.616 & 2.885 & 0.845 & 0.626 & 0.825 \\ 
			0.01 & 1 & 0.5 & rand & 0.085 & 0.145 & 0.988 & 0.578 & 3.296 & 0.854 & 0.583 & 0.864 \\ 
			0.01 & 1 & 1 & clus & 0.120 & 0.248 & 0.275 & 0.533 & 0.591 & 0.944 & 0.591 & 0.594 \\ 
			0.01 & 1 & 1 & rand & 0.077 & 0.117 & 0.148 & 0.519 & 0.622 & 0.945 & 0.609 & 0.655 \\ 
			0.01 & 1 & 1.5 & clus & 0.092 & 0.230 & 0.159 & 0.550 & 0.580 & 0.946 & 0.600 & 0.667 \\ 
			0.01 & 1 & 1.5 & rand & 0.073 & 0.109 & 0.086 & 0.537 & 0.558 & 0.944 & 0.634 & 0.696 \\ 
			0.01 & 1 & 2 & clus & 0.067 & 0.224 & 0.197 & 0.610 & 0.659 & 0.956 & 0.607 & 0.638 \\ 
			0.01 & 1 & 2 & rand & 0.060 & 0.113 & 0.102 & 0.606 & 0.632 & 0.944 & 0.632 & 0.684 \\ 
			0.01 & 10 & 0.5 & clus & 0.509 & 1.420 & 2.365 & 0.494 & 0.497 & 0.774 & 0.470 & 0.344 \\ 
			0.01 & 10 & 0.5 & rand & 0.438 & 1.156 & 1.594 & 0.451 & 0.418 & 0.769 & 0.408 & 0.332 \\ 
			0.01 & 10 & 1 & clus & 0.604 & 1.436 & 1.138 & 0.331 & 0.457 & 0.944 & 0.436 & 0.495 \\ 
			0.01 & 10 & 1 & rand & 0.502 & 1.094 & 0.690 & 0.300 & 0.420 & 0.942 & 0.417 & 0.526 \\ 
			0.01 & 10 & 1.5 & clus & 0.490 & 1.456 & 1.023 & 0.371 & 0.507 & 0.954 & 0.444 & 0.527 \\ 
			0.01 & 10 & 1.5 & rand & 0.459 & 1.090 & 0.662 & 0.316 & 0.489 & 0.939 & 0.422 & 0.559 \\ 
			0.01 & 10 & 2 & clus & 0.409 & 1.462 & 1.079 & 0.406 & 0.542 & 0.953 & 0.452 & 0.514 \\ 
			0.01 & 10 & 2 & rand & 0.392 & 1.034 & 0.676 & 0.374 & 0.554 & 0.940 & 0.454 & 0.560 \\ 
			0.01 & 100 & 0.5 & clus & 1.442 & 8.372 & 7.465 & 0.556 & 0.623 & 0.746 & 0.382 & 0.431 \\ 
			0.01 & 100 & 0.5 & rand & 1.446 & 9.610 & 5.844 & 0.558 & 0.497 & 0.748 & 0.354 & 0.419 \\ 
			0.01 & 100 & 1 & clus & 2.207 & 8.448 & 6.252 & 0.300 & 0.415 & 0.952 & 0.370 & 0.448 \\ 
			0.01 & 100 & 1 & rand & 2.230 & 9.378 & 4.935 & 0.308 & 0.398 & 0.939 & 0.365 & 0.464 \\ 
			0.01 & 100 & 1.5 & clus & 2.438 & 8.405 & 6.458 & 0.291 & 0.421 & 0.952 & 0.372 & 0.456 \\ 
			0.01 & 100 & 1.5 & rand & 2.436 & 9.529 & 5.015 & 0.294 & 0.396 & 0.940 & 0.369 & 0.472 \\ 
			0.01 & 100 & 2 & clus & 2.400 & 8.565 & 6.592 & 0.290 & 0.418 & 0.949 & 0.370 & 0.456 \\ 
			0.01 & 100 & 2 & rand & 2.417 & 9.592 & 5.044 & 0.287 & 0.396 & 0.945 & 0.364 & 0.483 \\ 
			\hline
			0.1 & 1 & 0.5 & clus & 0.269 & 0.523 & 1.732 & 0.579 & 2.041 & 0.940 & 0.658 & 0.858 \\ 
			0.1 & 1 & 0.5 & rand & 0.174 & 0.223 & 1.026 & 0.534 & 2.203 & 0.940 & 0.631 & 0.901 \\ 
			0.1 & 1 & 1 & clus & 0.216 & 0.362 & 0.382 & 0.597 & 0.662 & 0.954 & 0.664 & 0.684 \\ 
			0.1 & 1 & 1 & rand & 0.175 & 0.219 & 0.243 & 0.559 & 0.702 & 0.949 & 0.664 & 0.719 \\ 
			0.1 & 1 & 1.5 & clus & 0.193 & 0.329 & 0.265 & 0.616 & 0.597 & 0.951 & 0.674 & 0.706 \\ 
			0.1 & 1 & 1.5 & rand & 0.167 & 0.206 & 0.184 & 0.580 & 0.578 & 0.948 & 0.681 & 0.710 \\ 
			0.1 & 1 & 2 & clus & 0.161 & 0.335 & 0.304 & 0.626 & 0.635 & 0.952 & 0.675 & 0.698 \\ 
			0.1 & 1 & 2 & rand & 0.156 & 0.210 & 0.198 & 0.583 & 0.618 & 0.948 & 0.676 & 0.716 \\ 
			0.1 & 10 & 0.5 & clus & 0.620 & 1.551 & 2.442 & 0.626 & 0.811 & 0.870 & 0.624 & 0.627 \\ 
			0.1 & 10 & 0.5 & rand & 0.536 & 1.259 & 1.680 & 0.592 & 0.711 & 0.868 & 0.574 & 0.632 \\ 
			0.1 & 10 & 1 & clus & 0.686 & 1.573 & 1.227 & 0.504 & 0.594 & 0.949 & 0.621 & 0.670 \\ 
			0.1 & 10 & 1 & rand & 0.607 & 1.209 & 0.785 & 0.474 & 0.567 & 0.944 & 0.590 & 0.692 \\ 
			0.1 & 10 & 1.5 & clus & 0.570 & 1.529 & 1.126 & 0.524 & 0.654 & 0.953 & 0.615 & 0.688 \\ 
			0.1 & 10 & 1.5 & rand & 0.549 & 1.176 & 0.775 & 0.482 & 0.628 & 0.942 & 0.596 & 0.707 \\ 
			0.1 & 10 & 2 & clus & 0.501 & 1.543 & 1.178 & 0.564 & 0.698 & 0.956 & 0.624 & 0.686 \\ 
			0.1 & 10 & 2 & rand & 0.490 & 1.121 & 0.764 & 0.528 & 0.674 & 0.942 & 0.617 & 0.711 \\ 
			0.1 & 100 & 0.5 & clus & 1.496 & 8.350 & 7.602 & 0.624 & 0.655 & 0.802 & 0.440 & 0.464 \\ 
			0.1 & 100 & 0.5 & rand & 1.552 & 9.830 & 5.944 & 0.615 & 0.540 & 0.802 & 0.412 & 0.463 \\ 
			0.1 & 100 & 1 & clus & 2.331 & 8.652 & 6.363 & 0.348 & 0.449 & 0.952 & 0.437 & 0.498 \\ 
			0.1 & 100 & 1 & rand & 2.300 & 9.425 & 5.007 & 0.350 & 0.431 & 0.939 & 0.418 & 0.511 \\ 
			0.1 & 100 & 1.5 & clus & 2.529 & 8.526 & 6.600 & 0.333 & 0.448 & 0.952 & 0.430 & 0.504 \\ 
			0.1 & 100 & 1.5 & rand & 2.558 & 9.578 & 5.129 & 0.332 & 0.424 & 0.941 & 0.416 & 0.519 \\ 
			0.1 & 100 & 2 & clus & 2.497 & 8.721 & 6.707 & 0.330 & 0.445 & 0.950 & 0.434 & 0.505 \\ 
			0.1 & 100 & 2 & rand & 2.478 & 9.667 & 5.105 & 0.331 & 0.432 & 0.944 & 0.416 & 0.534 \\ 
			\hline
			1 & 1 & 0.5 & clus & 1.248 & 1.574 & 2.655 & 0.885 & 1.163 & 0.928 & 0.869 & 0.917 \\ 
			1 & 1 & 0.5 & rand & 1.134 & 1.247 & 2.010 & 0.839 & 1.218 & 0.927 & 0.854 & 0.929 \\ 
			1 & 1 & 1 & clus & 1.210 & 1.467 & 1.389 & 0.861 & 0.908 & 0.925 & 0.865 & 0.878 \\ 
			1 & 1 & 1 & rand & 1.143 & 1.221 & 1.211 & 0.841 & 0.910 & 0.922 & 0.866 & 0.896 \\ 
			1 & 1 & 1.5 & clus & 1.154 & 1.417 & 1.281 & 0.865 & 0.891 & 0.923 & 0.869 & 0.890 \\ 
			1 & 1 & 1.5 & rand & 1.122 & 1.217 & 1.160 & 0.844 & 0.872 & 0.922 & 0.870 & 0.890 \\ 
			1 & 1 & 2 & clus & 1.119 & 1.415 & 1.338 & 0.872 & 0.908 & 0.923 & 0.866 & 0.884 \\ 
			1 & 1 & 2 & rand & 1.106 & 1.200 & 1.165 & 0.861 & 0.893 & 0.922 & 0.871 & 0.894 \\ 
			1 & 10 & 0.5 & clus & 1.541 & 2.604 & 3.417 & 0.645 & 0.978 & 0.946 & 0.686 & 0.760 \\ 
			1 & 10 & 0.5 & rand & 1.464 & 2.273 & 2.634 & 0.579 & 0.917 & 0.946 & 0.648 & 0.758 \\ 
			1 & 10 & 1 & clus & 1.621 & 2.639 & 2.229 & 0.636 & 0.684 & 0.952 & 0.700 & 0.733 \\ 
			1 & 10 & 1 & rand & 1.528 & 2.197 & 1.738 & 0.572 & 0.632 & 0.948 & 0.667 & 0.732 \\ 
			1 & 10 & 1.5 & clus & 1.530 & 2.561 & 2.084 & 0.664 & 0.693 & 0.952 & 0.703 & 0.740 \\ 
			1 & 10 & 1.5 & rand & 1.488 & 2.249 & 1.720 & 0.588 & 0.634 & 0.948 & 0.669 & 0.731 \\ 
			1 & 10 & 2 & clus & 1.436 & 2.660 & 2.159 & 0.671 & 0.706 & 0.953 & 0.703 & 0.740 \\ 
			1 & 10 & 2 & rand & 1.428 & 2.125 & 1.697 & 0.606 & 0.647 & 0.947 & 0.679 & 0.736 \\ 
			1 & 100 & 0.5 & clus & 2.416 & 9.544 & 8.559 & 0.800 & 0.852 & 0.918 & 0.638 & 0.676 \\ 
			1 & 100 & 0.5 & rand & 2.471 & 10.861 & 6.855 & 0.779 & 0.766 & 0.914 & 0.618 & 0.681 \\ 
			1 & 100 & 1 & clus & 3.231 & 9.791 & 7.529 & 0.583 & 0.664 & 0.951 & 0.641 & 0.704 \\ 
			1 & 100 & 1 & rand & 3.186 & 10.233 & 5.884 & 0.581 & 0.629 & 0.945 & 0.628 & 0.720 \\ 
			1 & 100 & 1.5 & clus & 3.464 & 9.666 & 7.374 & 0.562 & 0.645 & 0.952 & 0.639 & 0.707 \\ 
			1 & 100 & 1.5 & rand & 3.477 & 10.522 & 6.014 & 0.556 & 0.624 & 0.944 & 0.628 & 0.723 \\ 
			1 & 100 & 2 & clus & 3.338 & 9.568 & 7.566 & 0.555 & 0.649 & 0.951 & 0.641 & 0.710 \\ 
			1 & 100 & 2 & rand & 3.388 & 10.500 & 6.018 & 0.555 & 0.622 & 0.946 & 0.624 & 0.726 \\ 
			\hline
		\end{tabular}
\end{table}

\begin{table}
	\caption{\label{pred_WP} Median predictions results, comparing mean square prediction error (MSPE), ratio of 
		median estimated versus true KSE, and prediction interval coverage (PIC) for a positive Wishart covariance.}
	\scriptsize
		\begin{tabular}{llll|rrr|rr|rrr}
			\hline
			& & & & & MSPE & & rKSE & & & PIC & \\
			$\sigma_{\delta}^2$ & $\sigma_{\epsilon}^2$ & $b$ & design & True & AECM & EM & AECM & EM & True & AECM & EM \\
			\hline
			0.01 & 1 & 0.5 & clus & 0.477 & 1.024 & 2.989 & 0.697 & 2.398 & 0.856 & 0.586 & 0.740 \\ 
			0.01 & 1 & 0.5 & rand & 0.067 & 0.110 & 1.245 & 0.596 & 3.971 & 0.857 & 0.546 & 0.909 \\ 
			0.01 & 1 & 1 & clus & 0.119 & 0.217 & 0.653 & 0.502 & 0.671 & 0.950 & 0.599 & 0.555 \\ 
			0.01 & 1 & 1 & rand & 0.079 & 0.114 & 0.324 & 0.529 & 1.446 & 0.947 & 0.647 & 0.734 \\ 
			0.01 & 1 & 1.5 & clus & 0.114 & 0.211 & 0.160 & 0.490 & 0.500 & 0.951 & 0.597 & 0.664 \\ 
			0.01 & 1 & 1.5 & rand & 0.077 & 0.114 & 0.091 & 0.563 & 0.575 & 0.954 & 0.696 & 0.726 \\ 
			0.01 & 1 & 2 & clus & 0.067 & 0.209 & 0.203 & 0.601 & 0.633 & 0.957 & 0.595 & 0.615 \\ 
			0.01 & 1 & 2 & rand & 0.063 & 0.118 & 0.121 & 0.582 & 0.629 & 0.952 & 0.690 & 0.717 \\ 
			0.01 & 10 & 0.5 & clus & 1.022 & 3.013 & 3.673 & 0.526 & 0.520 & 0.781 & 0.394 & 0.345 \\ 
			0.01 & 10 & 0.5 & rand & 0.551 & 1.379 & 2.126 & 0.474 & 0.513 & 0.784 & 0.397 & 0.316 \\ 
			0.01 & 10 & 1 & clus & 0.840 & 1.989 & 1.768 & 0.371 & 0.383 & 0.932 & 0.440 & 0.454 \\ 
			0.01 & 10 & 1 & rand & 0.568 & 1.092 & 1.035 & 0.372 & 0.361 & 0.937 & 0.457 & 0.434 \\ 
			0.01 & 10 & 1.5 & clus & 0.724 & 2.003 & 1.142 & 0.424 & 0.420 & 0.937 & 0.456 & 0.512 \\ 
			0.01 & 10 & 1.5 & rand & 0.510 & 1.016 & 0.740 & 0.393 & 0.393 & 0.946 & 0.486 & 0.524 \\ 
			0.01 & 10 & 2 & clus & 0.448 & 1.975 & 1.145 & 0.483 & 0.467 & 0.954 & 0.462 & 0.531 \\ 
			0.01 & 10 & 2 & rand & 0.403 & 1.072 & 0.770 & 0.451 & 0.442 & 0.954 & 0.479 & 0.536 \\ 
			0.01 & 100 & 0.5 & clus & 2.711 & 14.125 & 10.203 & 0.588 & 0.645 & 0.757 & 0.352 & 0.460 \\ 
			0.01 & 100 & 0.5 & rand & 2.220 & 11.455 & 7.379 & 0.487 & 0.615 & 0.762 & 0.319 & 0.408 \\ 
			0.01 & 100 & 1 & clus & 3.653 & 13.494 & 8.433 & 0.370 & 0.507 & 0.890 & 0.382 & 0.473 \\ 
			0.01 & 100 & 1 & rand & 3.044 & 10.525 & 6.489 & 0.355 & 0.476 & 0.889 & 0.367 & 0.450 \\ 
			0.01 & 100 & 1.5 & clus & 3.163 & 13.526 & 7.551 & 0.356 & 0.476 & 0.912 & 0.378 & 0.490 \\ 
			0.01 & 100 & 1.5 & rand & 2.805 & 10.097 & 5.913 & 0.328 & 0.443 & 0.913 & 0.375 & 0.448 \\ 
			0.01 & 100 & 2 & clus & 2.563 & 12.677 & 7.467 & 0.344 & 0.495 & 0.926 & 0.390 & 0.502 \\ 
			0.01 & 100 & 2 & rand & 2.448 & 10.370 & 5.930 & 0.308 & 0.409 & 0.933 & 0.371 & 0.471 \\ 
			\hline
			0.1 & 1 & 0.5 & clus & 0.617 & 1.351 & 3.179 & 0.574 & 1.714 & 0.945 & 0.633 & 0.794 \\ 
			0.1 & 1 & 0.5 & rand & 0.178 & 0.247 & 1.462 & 0.570 & 2.329 & 0.944 & 0.645 & 0.933 \\ 
			0.1 & 1 & 1 & clus & 0.217 & 0.344 & 0.755 & 0.597 & 0.888 & 0.953 & 0.668 & 0.669 \\ 
			0.1 & 1 & 1 & rand & 0.176 & 0.211 & 0.416 & 0.624 & 1.205 & 0.952 & 0.708 & 0.804 \\ 
			0.1 & 1 & 1.5 & clus & 0.213 & 0.316 & 0.256 & 0.588 & 0.575 & 0.954 & 0.675 & 0.709 \\ 
			0.1 & 1 & 1.5 & rand & 0.176 & 0.214 & 0.190 & 0.626 & 0.612 & 0.952 & 0.723 & 0.741 \\ 
			0.1 & 1 & 2 & clus & 0.161 & 0.324 & 0.318 & 0.625 & 0.657 & 0.955 & 0.661 & 0.685 \\ 
			0.1 & 1 & 2 & rand & 0.159 & 0.217 & 0.216 & 0.628 & 0.720 & 0.952 & 0.716 & 0.751 \\ 
			0.1 & 10 & 0.5 & clus & 1.207 & 3.339 & 3.750 & 0.645 & 0.748 & 0.876 & 0.585 & 0.615 \\ 
			0.1 & 10 & 0.5 & rand & 0.623 & 1.469 & 2.213 & 0.623 & 0.831 & 0.882 & 0.582 & 0.672 \\ 
			0.1 & 10 & 1 & clus & 0.904 & 2.266 & 1.837 & 0.525 & 0.578 & 0.937 & 0.590 & 0.627 \\ 
			0.1 & 10 & 1 & rand & 0.661 & 1.202 & 1.143 & 0.560 & 0.592 & 0.942 & 0.620 & 0.654 \\ 
			0.1 & 10 & 1.5 & clus & 0.807 & 2.149 & 1.253 & 0.571 & 0.567 & 0.942 & 0.597 & 0.664 \\ 
			0.1 & 10 & 1.5 & rand & 0.617 & 1.106 & 0.833 & 0.561 & 0.560 & 0.949 & 0.660 & 0.706 \\ 
			0.1 & 10 & 2 & clus & 0.537 & 2.060 & 1.240 & 0.610 & 0.620 & 0.957 & 0.606 & 0.676 \\ 
			0.1 & 10 & 2 & rand & 0.501 & 1.157 & 0.862 & 0.595 & 0.609 & 0.956 & 0.656 & 0.720 \\ 
			0.1 & 100 & 0.5 & clus & 2.858 & 13.896 & 10.354 & 0.653 & 0.683 & 0.813 & 0.425 & 0.482 \\ 
			0.1 & 100 & 0.5 & rand & 2.297 & 11.898 & 7.632 & 0.575 & 0.677 & 0.823 & 0.386 & 0.450 \\ 
			0.1 & 100 & 1 & clus & 3.707 & 13.924 & 8.734 & 0.408 & 0.523 & 0.895 & 0.438 & 0.508 \\ 
			0.1 & 100 & 1 & rand & 2.980 & 10.811 & 6.502 & 0.408 & 0.496 & 0.892 & 0.425 & 0.492 \\ 
			0.1 & 100 & 1.5 & clus & 3.235 & 13.385 & 7.728 & 0.401 & 0.508 & 0.916 & 0.438 & 0.532 \\ 
			0.1 & 100 & 1.5 & rand & 2.916 & 10.334 & 6.012 & 0.377 & 0.476 & 0.916 & 0.420 & 0.500 \\ 
			0.1 & 100 & 2 & clus & 2.669 & 12.698 & 7.707 & 0.380 & 0.511 & 0.931 & 0.448 & 0.542 \\ 
			0.1 & 100 & 2 & rand & 2.569 & 10.449 & 5.963 & 0.355 & 0.441 & 0.931 & 0.430 & 0.514 \\ 
			\hline
			1 & 1 & 0.5 & clus & 1.575 & 2.354 & 4.084 & 0.870 & 1.079 & 0.930 & 0.864 & 0.897 \\ 
			1 & 1 & 0.5 & rand & 1.145 & 1.264 & 2.466 & 0.856 & 1.260 & 0.930 & 0.861 & 0.938 \\ 
			1 & 1 & 1 & clus & 1.240 & 1.459 & 1.773 & 0.851 & 0.945 & 0.925 & 0.860 & 0.870 \\ 
			1 & 1 & 1 & rand & 1.144 & 1.223 & 1.400 & 0.861 & 0.998 & 0.924 & 0.873 & 0.906 \\ 
			1 & 1 & 1.5 & clus & 1.218 & 1.411 & 1.281 & 0.854 & 0.875 & 0.923 & 0.866 & 0.889 \\ 
			1 & 1 & 1.5 & rand & 1.134 & 1.206 & 1.168 & 0.860 & 0.877 & 0.923 & 0.877 & 0.893 \\ 
			1 & 1 & 2 & clus & 1.114 & 1.412 & 1.328 & 0.871 & 0.900 & 0.923 & 0.867 & 0.887 \\ 
			1 & 1 & 2 & rand & 1.109 & 1.221 & 1.194 & 0.865 & 0.907 & 0.925 & 0.878 & 0.898 \\ 
			1 & 10 & 0.5 & clus & 2.059 & 4.285 & 4.822 & 0.639 & 0.868 & 0.951 & 0.671 & 0.738 \\ 
			1 & 10 & 0.5 & rand & 1.581 & 2.482 & 3.186 & 0.588 & 1.056 & 0.950 & 0.660 & 0.790 \\ 
			1 & 10 & 1 & clus & 1.924 & 3.353 & 2.844 & 0.630 & 0.671 & 0.952 & 0.686 & 0.713 \\ 
			1 & 10 & 1 & rand & 1.595 & 2.223 & 2.111 & 0.623 & 0.747 & 0.951 & 0.694 & 0.752 \\ 
			1 & 10 & 1.5 & clus & 1.741 & 3.235 & 2.255 & 0.623 & 0.635 & 0.953 & 0.686 & 0.726 \\ 
			1 & 10 & 1.5 & rand & 1.545 & 2.087 & 1.820 & 0.648 & 0.654 & 0.954 & 0.715 & 0.748 \\ 
			1 & 10 & 2 & clus & 1.471 & 3.164 & 2.228 & 0.667 & 0.680 & 0.953 & 0.693 & 0.730 \\ 
			1 & 10 & 2 & rand & 1.442 & 2.174 & 1.819 & 0.663 & 0.680 & 0.953 & 0.716 & 0.757 \\ 
			1 & 100 & 0.5 & clus & 3.724 & 14.839 & 10.887 & 0.804 & 0.859 & 0.928 & 0.624 & 0.666 \\ 
			1 & 100 & 0.5 & rand & 3.213 & 12.601 & 8.405 & 0.768 & 0.881 & 0.929 & 0.615 & 0.677 \\ 
			1 & 100 & 1 & clus & 4.561 & 14.698 & 9.633 & 0.635 & 0.701 & 0.920 & 0.631 & 0.694 \\ 
			1 & 100 & 1 & rand & 3.950 & 11.737 & 7.406 & 0.642 & 0.695 & 0.919 & 0.624 & 0.702 \\ 
			1 & 100 & 1.5 & clus & 4.211 & 14.104 & 8.698 & 0.647 & 0.695 & 0.931 & 0.622 & 0.685 \\ 
			1 & 100 & 1.5 & rand & 3.803 & 11.181 & 6.954 & 0.638 & 0.693 & 0.932 & 0.628 & 0.710 \\ 
			1 & 100 & 2 & clus & 3.634 & 13.968 & 8.757 & 0.624 & 0.696 & 0.940 & 0.636 & 0.705 \\ 
			1 & 100 & 2 & rand & 3.567 & 11.602 & 7.010 & 0.610 & 0.659 & 0.942 & 0.637 & 0.719 \\ 
			\hline
		\end{tabular}
\end{table}

\begin{table}
	\caption{\label{pred_WN} Median predictions results, comparing mean square prediction error (MSPE), ratio of 
		median estimated versus true KSE, and prediction interval coverage (PIC) for a Wishart covariance.}
	\scriptsize
		\begin{tabular}{llll|rrr|rr|rrr}
			\hline
			& & & & & MSPE & & rKSE & & & PIC & \\
			$\sigma_{\delta}^2$ & $\sigma_{\epsilon}^2$ & $b$ & design & True & AECM & EM & AECM & EM & True & AECM & EM \\
			\hline
			0.01 & 1 & 0.5 & clus & 0.586 & 1.579 & 3.139 & 0.689 & 2.380 & 0.856 & 0.576 & 0.709 \\ 
			0.01 & 1 & 0.5 & rand & 0.082 & 0.140 & 1.342 & 0.645 & 3.727 & 0.854 & 0.568 & 0.907 \\ 
			0.01 & 1 & 1 & clus & 0.123 & 0.236 & 0.654 & 0.499 & 0.662 & 0.947 & 0.591 & 0.560 \\ 
			0.01 & 1 & 1 & rand & 0.080 & 0.110 & 0.313 & 0.530 & 1.353 & 0.947 & 0.664 & 0.732 \\ 
			0.01 & 1 & 1.5 & clus & 0.114 & 0.219 & 0.169 & 0.481 & 0.493 & 0.954 & 0.605 & 0.655 \\ 
			0.01 & 1 & 1.5 & rand & 0.079 & 0.109 & 0.095 & 0.568 & 0.579 & 0.958 & 0.694 & 0.745 \\ 
			0.01 & 1 & 2 & clus & 0.069 & 0.219 & 0.218 & 0.587 & 0.611 & 0.954 & 0.587 & 0.626 \\ 
			0.01 & 1 & 2 & rand & 0.067 & 0.119 & 0.117 & 0.565 & 0.608 & 0.953 & 0.680 & 0.711 \\ 
			0.01 & 10 & 0.5 & clus & 1.093 & 2.790 & 3.987 & 0.503 & 0.529 & 0.784 & 0.394 & 0.350 \\ 
			0.01 & 10 & 0.5 & rand & 0.546 & 1.332 & 2.121 & 0.463 & 0.501 & 0.784 & 0.387 & 0.320 \\ 
			0.01 & 10 & 1 & clus & 0.855 & 1.979 & 1.665 & 0.352 & 0.364 & 0.941 & 0.424 & 0.467 \\ 
			0.01 & 10 & 1 & rand & 0.587 & 1.152 & 1.040 & 0.333 & 0.340 & 0.943 & 0.432 & 0.442 \\ 
			0.01 & 10 & 1.5 & clus & 0.719 & 1.831 & 1.114 & 0.390 & 0.396 & 0.943 & 0.444 & 0.508 \\ 
			0.01 & 10 & 1.5 & rand & 0.543 & 1.074 & 0.734 & 0.380 & 0.382 & 0.955 & 0.454 & 0.518 \\ 
			0.01 & 10 & 2 & clus & 0.478 & 1.794 & 1.140 & 0.463 & 0.440 & 0.958 & 0.454 & 0.529 \\ 
			0.01 & 10 & 2 & rand & 0.442 & 1.131 & 0.755 & 0.430 & 0.430 & 0.954 & 0.462 & 0.528 \\ 
			0.01 & 100 & 0.5 & clus & 2.681 & 13.242 & 10.101 & 0.593 & 0.639 & 0.758 & 0.336 & 0.444 \\ 
			0.01 & 100 & 0.5 & rand & 2.204 & 11.911 & 7.389 & 0.514 & 0.675 & 0.769 & 0.318 & 0.424 \\ 
			0.01 & 100 & 1 & clus & 3.582 & 13.106 & 8.480 & 0.348 & 0.503 & 0.888 & 0.370 & 0.471 \\ 
			0.01 & 100 & 1 & rand & 2.957 & 10.930 & 6.383 & 0.327 & 0.462 & 0.888 & 0.346 & 0.448 \\ 
			0.01 & 100 & 1.5 & clus & 3.216 & 11.952 & 7.602 & 0.348 & 0.501 & 0.908 & 0.378 & 0.500 \\ 
			0.01 & 100 & 1.5 & rand & 2.751 & 10.207 & 5.883 & 0.322 & 0.456 & 0.914 & 0.368 & 0.452 \\ 
			0.01 & 100 & 2 & clus & 2.483 & 12.425 & 7.471 & 0.338 & 0.510 & 0.920 & 0.384 & 0.487 \\ 
			0.01 & 100 & 2 & rand & 2.385 & 10.409 & 5.858 & 0.328 & 0.436 & 0.926 & 0.370 & 0.458 \\ 
			\hline
			0.1 & 1 & 0.5 & clus & 0.684 & 1.390 & 3.265 & 0.612 & 1.725 & 0.940 & 0.643 & 0.786 \\ 
			0.1 & 1 & 0.5 & rand & 0.178 & 0.242 & 1.487 & 0.596 & 2.357 & 0.939 & 0.655 & 0.925 \\ 
			0.1 & 1 & 1 & clus & 0.214 & 0.344 & 0.739 & 0.603 & 0.916 & 0.954 & 0.674 & 0.679 \\ 
			0.1 & 1 & 1 & rand & 0.177 & 0.206 & 0.406 & 0.612 & 1.187 & 0.953 & 0.709 & 0.809 \\ 
			0.1 & 1 & 1.5 & clus & 0.213 & 0.326 & 0.264 & 0.574 & 0.574 & 0.954 & 0.670 & 0.697 \\ 
			0.1 & 1 & 1.5 & rand & 0.178 & 0.214 & 0.192 & 0.623 & 0.617 & 0.953 & 0.720 & 0.743 \\ 
			0.1 & 1 & 2 & clus & 0.164 & 0.318 & 0.323 & 0.627 & 0.661 & 0.954 & 0.668 & 0.700 \\ 
			0.1 & 1 & 2 & rand & 0.161 & 0.222 & 0.216 & 0.611 & 0.705 & 0.952 & 0.717 & 0.751 \\ 
			0.1 & 10 & 0.5 & clus & 1.166 & 2.998 & 4.082 & 0.650 & 0.744 & 0.870 & 0.586 & 0.616 \\ 
			0.1 & 10 & 0.5 & rand & 0.617 & 1.448 & 2.253 & 0.621 & 0.811 & 0.880 & 0.587 & 0.664 \\ 
			0.1 & 10 & 1 & clus & 0.951 & 2.085 & 1.752 & 0.504 & 0.548 & 0.946 & 0.583 & 0.643 \\ 
			0.1 & 10 & 1 & rand & 0.666 & 1.240 & 1.131 & 0.515 & 0.536 & 0.947 & 0.607 & 0.657 \\ 
			0.1 & 10 & 1.5 & clus & 0.826 & 1.959 & 1.216 & 0.527 & 0.543 & 0.946 & 0.586 & 0.663 \\ 
			0.1 & 10 & 1.5 & rand & 0.636 & 1.178 & 0.854 & 0.546 & 0.540 & 0.956 & 0.647 & 0.702 \\ 
			0.1 & 10 & 2 & clus & 0.571 & 1.955 & 1.234 & 0.593 & 0.593 & 0.958 & 0.603 & 0.675 \\ 
			0.1 & 10 & 2 & rand & 0.530 & 1.237 & 0.849 & 0.576 & 0.591 & 0.955 & 0.644 & 0.714 \\ 
			0.1 & 100 & 0.5 & clus & 2.728 & 13.341 & 10.273 & 0.651 & 0.685 & 0.819 & 0.424 & 0.480 \\ 
			0.1 & 100 & 0.5 & rand & 2.282 & 11.977 & 7.594 & 0.582 & 0.695 & 0.822 & 0.392 & 0.454 \\ 
			0.1 & 100 & 1 & clus & 3.668 & 13.421 & 8.622 & 0.409 & 0.528 & 0.892 & 0.435 & 0.496 \\ 
			0.1 & 100 & 1 & rand & 3.052 & 10.900 & 6.405 & 0.384 & 0.498 & 0.893 & 0.412 & 0.490 \\ 
			0.1 & 100 & 1.5 & clus & 3.316 & 12.098 & 7.664 & 0.398 & 0.523 & 0.910 & 0.438 & 0.524 \\ 
			0.1 & 100 & 1.5 & rand & 2.889 & 10.246 & 5.987 & 0.375 & 0.485 & 0.916 & 0.418 & 0.486 \\ 
			0.1 & 100 & 2 & clus & 2.570 & 12.533 & 7.525 & 0.387 & 0.536 & 0.921 & 0.446 & 0.534 \\ 
			0.1 & 100 & 2 & rand & 2.492 & 10.449 & 5.943 & 0.370 & 0.468 & 0.927 & 0.426 & 0.506 \\ 
			\hline
			1 & 1 & 0.5 & clus & 1.642 & 2.447 & 4.321 & 0.867 & 1.078 & 0.928 & 0.858 & 0.888 \\ 
			1 & 1 & 0.5 & rand & 1.150 & 1.271 & 2.414 & 0.865 & 1.263 & 0.928 & 0.862 & 0.940 \\ 
			1 & 1 & 1 & clus & 1.229 & 1.474 & 1.756 & 0.859 & 0.949 & 0.925 & 0.867 & 0.880 \\ 
			1 & 1 & 1 & rand & 1.140 & 1.217 & 1.394 & 0.855 & 0.991 & 0.924 & 0.872 & 0.908 \\ 
			1 & 1 & 1.5 & clus & 1.204 & 1.391 & 1.271 & 0.858 & 0.881 & 0.924 & 0.866 & 0.890 \\ 
			1 & 1 & 1.5 & rand & 1.136 & 1.207 & 1.168 & 0.860 & 0.885 & 0.923 & 0.879 & 0.894 \\ 
			1 & 1 & 2 & clus & 1.115 & 1.437 & 1.331 & 0.867 & 0.900 & 0.923 & 0.864 & 0.885 \\ 
			1 & 1 & 2 & rand & 1.112 & 1.220 & 1.189 & 0.854 & 0.904 & 0.924 & 0.870 & 0.892 \\ 
			1 & 10 & 0.5 & clus & 2.121 & 4.254 & 5.054 & 0.629 & 0.869 & 0.951 & 0.662 & 0.742 \\ 
			1 & 10 & 0.5 & rand & 1.590 & 2.466 & 3.171 & 0.607 & 1.048 & 0.950 & 0.657 & 0.790 \\ 
			1 & 10 & 1 & clus & 1.942 & 3.254 & 2.783 & 0.621 & 0.668 & 0.952 & 0.681 & 0.719 \\ 
			1 & 10 & 1 & rand & 1.612 & 2.270 & 2.097 & 0.610 & 0.708 & 0.953 & 0.683 & 0.743 \\ 
			1 & 10 & 1.5 & clus & 1.775 & 3.035 & 2.169 & 0.629 & 0.640 & 0.953 & 0.683 & 0.730 \\ 
			1 & 10 & 1.5 & rand & 1.574 & 2.157 & 1.803 & 0.644 & 0.644 & 0.954 & 0.707 & 0.753 \\ 
			1 & 10 & 2 & clus & 1.500 & 2.981 & 2.227 & 0.661 & 0.664 & 0.954 & 0.690 & 0.724 \\ 
			1 & 10 & 2 & rand & 1.455 & 2.229 & 1.794 & 0.638 & 0.659 & 0.953 & 0.696 & 0.751 \\ 
			1 & 100 & 0.5 & clus & 3.692 & 14.274 & 11.203 & 0.794 & 0.861 & 0.926 & 0.624 & 0.661 \\ 
			1 & 100 & 0.5 & rand & 3.210 & 12.881 & 8.484 & 0.757 & 0.884 & 0.932 & 0.605 & 0.674 \\ 
			1 & 100 & 1 & clus & 4.592 & 14.293 & 9.487 & 0.642 & 0.701 & 0.918 & 0.632 & 0.688 \\ 
			1 & 100 & 1 & rand & 3.917 & 11.703 & 7.325 & 0.639 & 0.700 & 0.920 & 0.613 & 0.694 \\ 
			1 & 100 & 1.5 & clus & 4.216 & 12.840 & 8.468 & 0.640 & 0.701 & 0.927 & 0.618 & 0.693 \\ 
			1 & 100 & 1.5 & rand & 3.817 & 11.379 & 6.975 & 0.627 & 0.685 & 0.933 & 0.627 & 0.704 \\ 
			1 & 100 & 2 & clus & 3.484 & 13.311 & 8.493 & 0.636 & 0.707 & 0.936 & 0.640 & 0.712 \\ 
			1 & 100 & 2 & rand & 3.316 & 11.494 & 6.852 & 0.624 & 0.688 & 0.938 & 0.629 & 0.717 \\ 
			\hline
		\end{tabular}
\end{table}
\subsubsection{Application: Predicting temperatures over the contiguous United States}
\begin{figure}[htbp]
	\begin{center}
		\includegraphics[height=\textwidth]{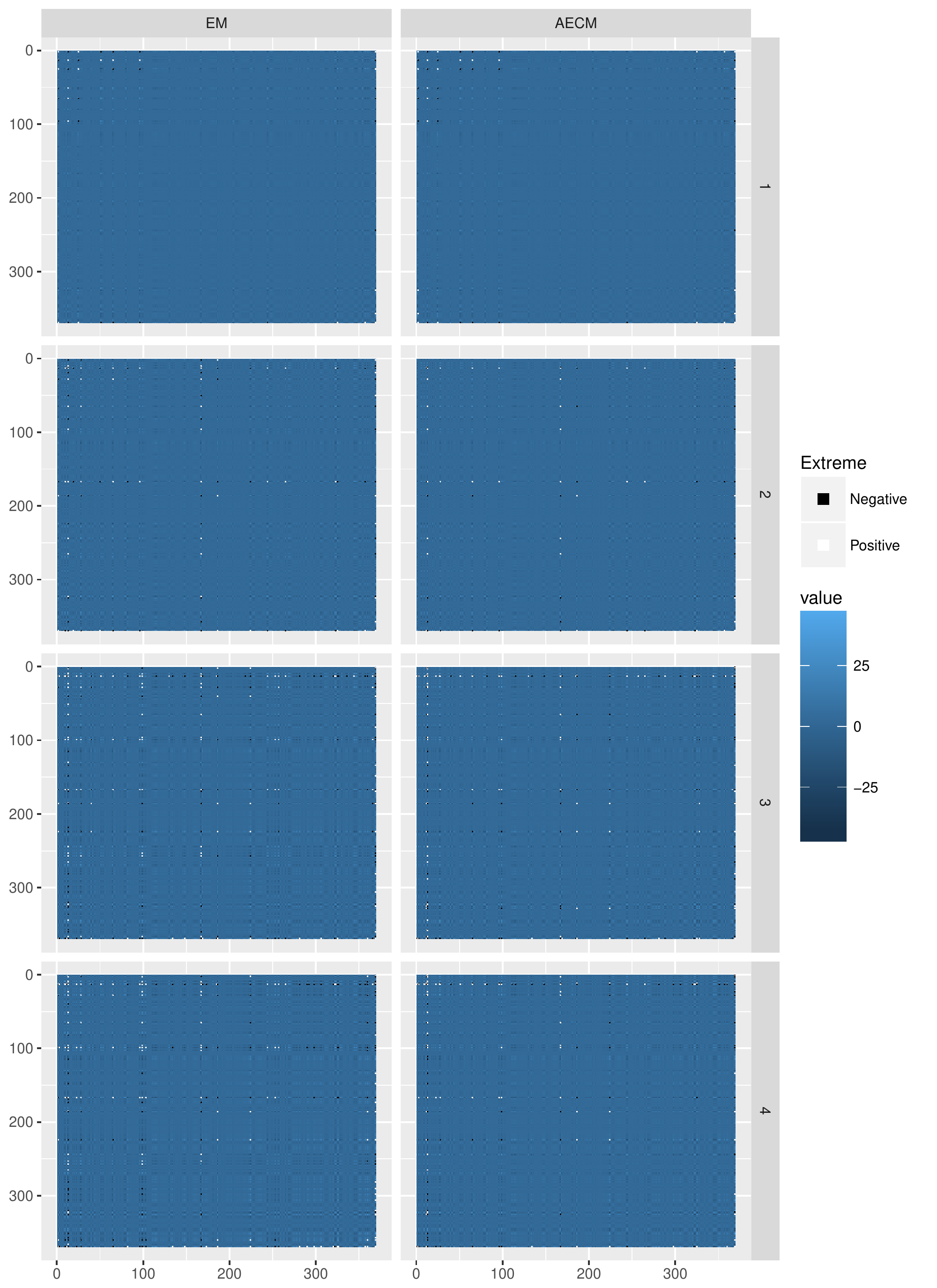}
	\end{center}
	\caption{$\hat{\bK}$ when assuming the full model. ``Extreme'' 
		values are defined as estimates beyond +/- 50.}
	\label{Kf}
\end{figure}
\begin{figure}[!htbp]
	\begin{center}
		\includegraphics[height=\textwidth]{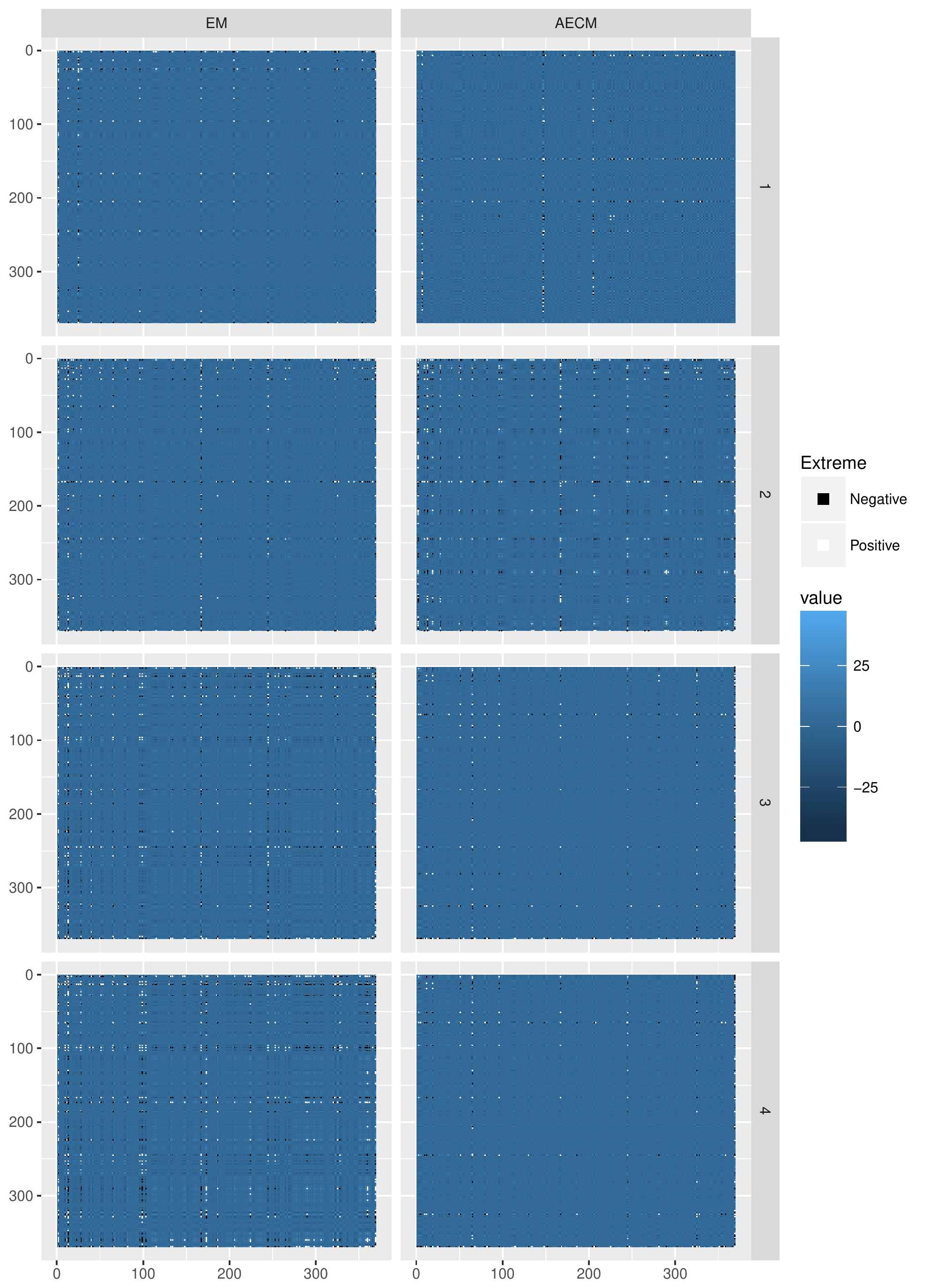}
	\end{center}
	\caption{$\hat{\bK}$ when assuming the reduced model. ``Extreme'' 
		values are defined as estimates beyond +/- 50.}
	\label{Kr}
\end{figure}

We applied our methodology to climatological data on temperature
recorded by the Cooperative Observer Program (COOP), archived 
at the US National Climate Data Center (NCDC) and available online at 
{\em http://www.image.ucar.edu/Data/US.monthly.met/}. As a response, we chose 
the average of mean monthly minimum and maximum temperature for April, 1990.
Complete image plots of $\hat{\bK}$ are provided in Figures \ref{Kf} and \ref{Kr} 
for the full and reduced models, respectively. In order to visually decipher smaller 
differences, all estimates of $\bK$ such that $\abs{\hat{K_{ij}}} > 50$ are truncated. 
These ``extreme'' values are then color-coded in black and white for emphasis.

\end{document}